%% file: Arxiv_Upload.tex
\documentclass{sig-alternate}

\usepackage{epstopdf}
\usepackage{chngcntr}
\usepackage{fancyhdr}

\usepackage{graphicx}
\usepackage{caption}
\usepackage{subcaption}
\usepackage{amsmath,amssymb}
\usepackage{cite}
\usepackage{bm,array}
\usepackage{color}
\usepackage{multicol}
\usepackage{algorithm}
\usepackage[noend]{algorithmic}

\usepackage{psfrag,mathrsfs,latexsym,graphics,graphicx,color,graphicx,caption}
\usepackage{verbatim}
\usepackage{url}
\usepackage[table]{xcolor}

\usepackage{blindtext}

\usepackage{etoolbox}
\makeatletter
\patchcmd{\maketitle}{\@copyrightspace}{}{}{}
\makeatother

\newcolumntype{C}{>{\centering}p{2.2em}}

\newtheorem{theorem}{Theorem}
\newtheorem{lemma}{Lemma}
\newtheorem{observation}{Observation}

\newtheorem{corollary}{Corollary}

\newtheorem{definition}{Definition}

\newtheorem{claim}{Claim}
\newtheorem{problem}{Problem}

\newcommand{\opnorm}[1]{{\left\vert\kern-0.22ex\left\vert\kern-0.22ex\left\vert #1
    \right\vert\kern-0.22ex\right\vert\kern-0.22ex\right\vert}}

\makeatletter
\def\bbordermatrix#1{\begingroup \m@th
  \@tempdima 4.75\p@
  \setbox\z@\vbox{%
    \def\cr{\crcr\noalign{\kern2\p@\global\let\cr\endline}}%
    \ialign{$##$\hfil\kern2\p@\kern\@tempdima&\thinspace\hfil$##$\hfil
      &&\quad\hfil$##$\hfil\crcr
      \omit\strut\hfil\crcr\noalign{\kern-\baselineskip}%
      #1\crcr\omit\strut\cr}}%
  \setbox\tw@\vbox{\unvcopy\z@\global\setbox\@ne\lastbox}%
  \setbox\tw@\hbox{\unhbox\@ne\unskip\global\setbox\@ne\lastbox}%
  \setbox\tw@\hbox{$\kern\wd\@ne\kern-\@tempdima\left[\kern-\wd\@ne
    \global\setbox\@ne\vbox{\box\@ne\kern2\p@}%
    \vcenter{\kern-\ht\@ne\unvbox\z@\kern-\baselineskip}\,\right]$}%
  \null\;\vbox{\kern\ht\@ne\box\tw@}\endgroup}
\makeatother

\newcolumntype{L}[1]{>{\raggedright\let\newline\\\arraybackslash\hspace{0pt}}m{#1}}
\newcolumntype{C}[1]{>{\centering\let\newline\\\arraybackslash\hspace{0pt}}m{#1}}
\newcolumntype{R}[1]{>{\raggedleft\let\newline\\\arraybackslash\hspace{0pt}}m{#1}}

\begin{document}

\thispagestyle{fancy}

\title{Cascading Failures in Power Grids -- \\Analysis and Algorithms}

\pagenumbering{arabic}

\date{}

\author{Saleh Soltan\\
       {Electrical Engineering}\\
       {Columbia University}\\
       {New York, NY}\\
       {saleh@ee.columbia.edu}
\and Dorian Mazauric\\
       {Laboratoire d'Informatique }\\
      {Fondamentale de Marseille}\\
      {Marseille, France}\\
       {dorian.mazauric@lif.univ-mrs.fr}
\and Gil Zussman\\
       {Electrical Engineering}\\
       {Columbia University}\\
       {New York, NY}\\
       {gil@ee.columbia.edu}
}

\maketitle

\begin{abstract}
This paper focuses on \emph{cascading line failures in the transmission system of the power grid}.
Recent large-scale power outages demonstrated the limitations of percolation- and epid- emic-based tools in modeling cascades.
Hence, we study cascades by using computational tools and a linearized power flow model. We first obtain results regarding the Moore-Penrose pseudo-inverse of the power grid admittance matrix. Based on these results, we study the impact of a \emph{single line failure} on the flows on other lines. We also illustrate via simulation the impact of the distance and resistance distance on the flow increase following a failure, and discuss the difference from the epidemic models.  We then study the \emph{cascade properties}, considering metrics such as the distance between failures and the fraction of demand (load) satisfied after the cascade (yield).
We use the pseudo-inverse of admittance matrix to develop an efficient \emph{algorithm to identify the cascading failure evolution}, which can be a building block for cascade mitigation. Finally, we show that finding the set of lines whose removal has the most significant impact (under various metrics) is NP-Hard and introduce a simple heuristic for the minimum yield problem. Overall, the results demonstrate that using the resistance distance and the pseudo-inverse of admittance matrix provides important insights and can support the development of efficient algorithms.
\end{abstract}

\smallskip

\setlength{\textfloatsep}{0pt}
\input{Introduction}

\input{Related}

\input{Model}
\input{MatrixProperties}
\input{FailureAnalysis}

\input{Properties}

\input{LinAlgebra}

\input{Hardness}

\input{Conclusion}
\section*{Acknowledgement}
This work was supported in part by CIAN NSF ERC under grant EEC-0812072, NSF grant CNS-1018379, and DTRA grant HDTRA1-13-1-0021.

\scriptsize
\bibliographystyle{abbrv}
\bibliography{bib}

\normalsize

\appendix
\makeatletter
\counterwithin{theorem}{section}
\counterwithin{lemma}{section}
\counterwithin{definition}{section}
\counterwithin{corollary}{section}
\counterwithin{equation}{section}
\input{Appendix-LinearAlgebra}
\input{Appendix-Proofs}

\end{document}

%% file: Introduction.tex
\section{Introduction}
\label{section:introduction}
\vspace{0.1cm}

Recent failures of the power grid (such as the 2003 and 2012 blackouts in the Northeastern U.S.~\cite{Blackout} and in India~\cite{IndiaBlackout}) demonstrated that large-scale failures will have devastating effects on almost every aspect in modern life.
The grid is vulnerable to natural disasters, such as earthquakes, hurricanes, and solar flares as well as to terrorist and Electromagnetic Pulse (EMP) attacks~\cite{EMP_flare}. Moreover, large scale cascades can be initiated by sporadic events \cite{Blackout,IndiaBlackout,SD_briefing}.

\begin{figure}[t]
\centering
\includegraphics[scale=0.22]{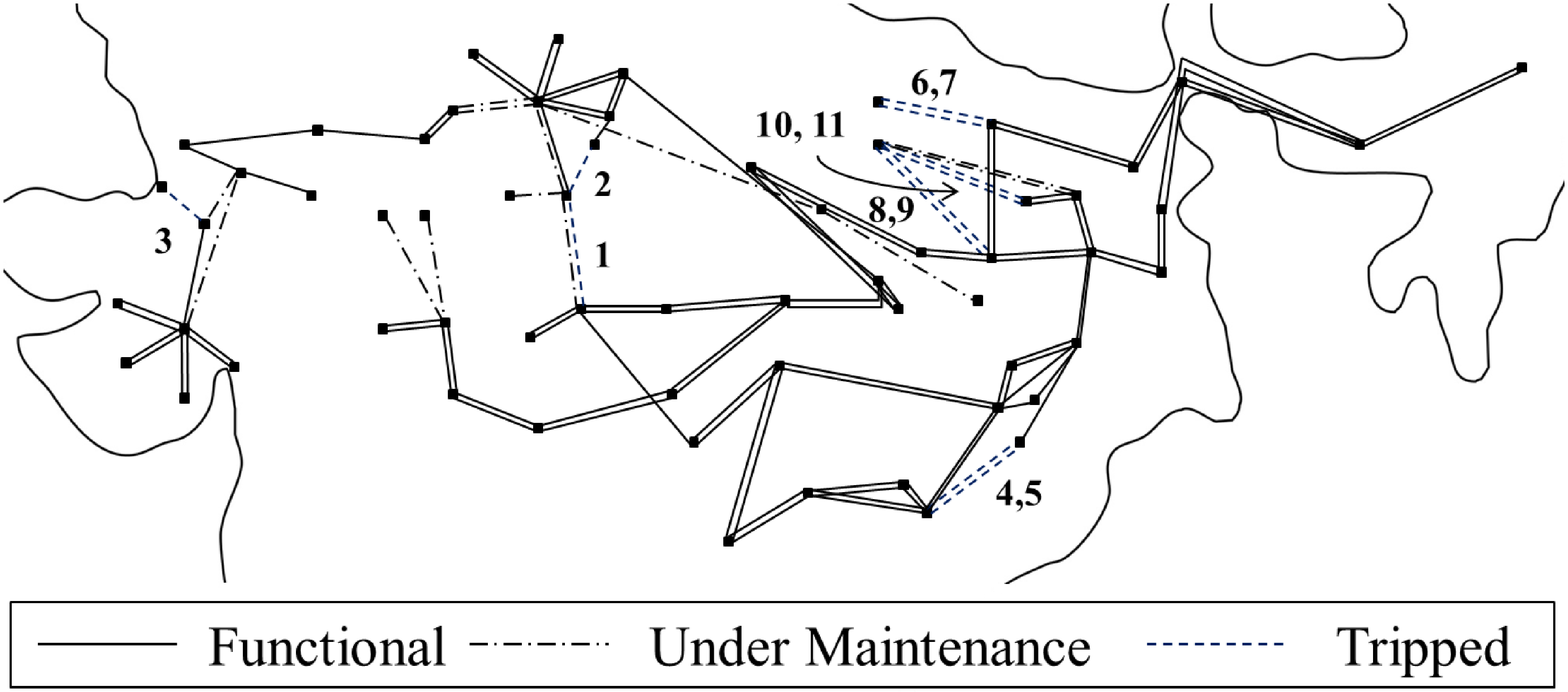}
\vspace{-0.4cm}
\caption{The first 11 line outages leading to the India blackout on July 2012~\cite{IndiaBlackout} (numbers show the order of outages).}
\vspace{0.1cm}
\label{figure:india}
\end{figure}

Therefore, there is a need to study the vulnerability of the power \emph{transmission} network.
Unlike graph-theoretical network flows, power flows are governed by the laws of physics and there are \emph{no strict capacity bounds on the lines}~\cite{bergenvittal}. Yet, there is a \emph{rating threshold} associated with each line -- if the flow exceeds the threshold, the line will eventually experience thermal failure.
Such an outage alters the network topology, giving rise to a different flow pattern which, in turn, could cause other line outages.  The repetition of this process constitutes a \emph{cascading failure} \cite{Chen2005318}.

Previous work (e.g., \cite{yeh,Chassin2005667,buldyrev2010catastrophic} and references therein) assumed that a line/node failure leads, with some probability,  to a failure of nearby nodes/lines. Such epidemic based modeling allows using percolation-based tools to analyze the cascade's effects. Yet, in real large scale cascades, a failure of a specific line can affect a remote line and \emph{the cascade does not necessarily develop in a contiguous manner}. For example, the evolution of the the cascade in India on July  2012 appears in Fig.~\ref{figure:india}.
Similar non-contiguous evolution was observed in a cascade in Southern California in 2011 \cite{SD_briefing,Bern2012ACM} and in simulation studies \cite{Bern2012ACM,SmartGridComm11rep}.

Motivated by this observation, we study the properties of the cascade and introduce algorithms to identify the cascading failure evolution and vulnerable lines. We employ the (linearized) \emph{direct-current (DC) power flow model},\footnote{The DC model is commonly used in large-scale contingency analysis of power grids \cite{Bie10,Dan2,pinar_power}.} which is a
 practical relaxation of the alternating-current (AC)  model, and the \emph{cascading failure model} of \cite{Dobson} (see also~\cite{Bie10,Dan2,SmartGridComm11rep,Bern2012ACM}).
Specifically, we first review the model and the Cascading Failure Evolution (CFE) Algorithm that has been used to identify the evolution of the cascade \cite{Chen2005318,Bie10,Dan2} (its complexity is $O(t|V|^3)$, where $|V|$ is the number of nodes and $t$ is the number of cascade rounds).

Then, in order to investigate the impact of a single edge failure on other edges, we use matrix analysis tools to study the properties of the admittance matrix of the grid\footnote{An $n\times n$ admittance matrix represents the admittance of the lines in a power grid with $n$ nodes.} and  \textit{Moore-Penrose Pseudo-inverse}~\cite{albert1972regression} of the admittance matrix. In particular, we provide a rank-1 update of the pseudo-inverse of the admittance matrix after a single edge failure.

We use these results along with the \emph{resistance distance} and \emph{Kirchhoff's index} notions\footnote{ These notions originate from Circuit Theory and are widely used in Chemistry~\cite{klein1993resistance}.} to study the impact of a \emph{single edge failure} on the flows on other edges. We obtain upper bounds on the flow changes after a single failure and study the robustness of specific graph classes. We also illustrate via simulations the relation between the flow changes after a failure and the distance (in hop count) and resistance distance from the failure in the U.S. Western interconnection as well as Erd\H{o}s-R\'{e}nyi~\cite{erdHos1959random}, Watts and Strogatz~\cite{watts1998collective}, and Bar\'{a}basi and Albert~\cite{barabasi1999emergence} graphs. These simulations show that there are cases in which an edge flow far away from the failure significantly increases. These \emph{average case} observations are clearly in contrast to the epidemic-based models.

We then consider the impact of a \emph{cascade}. We consider a few metrics: \emph{yield} (the fraction of demand satisfied after the cascade), \emph{number of line failures}, \emph{number of cascade rounds}, and the \emph{distance between consecutive failures}.
We generalize the results of~\cite{SmartGridComm11rep} and show that in the \emph{worst cases}, an initial single line failure may have severe effects while any super-set of  failures that includes that line have minor effects.
We then show that the metric values may be arbitrarily large or small (in case of the yield) even for a single initial line failure. We also prove that cascading failures may happen within arbitrarily long distance of each other and can last a large number of rounds. \emph{These characteristics are significantly different from those of the epidemic-based models}.
Finally, we show that a minor parameter change may have a significant impact.

Once lines fail, there is a
need for low complexity algorithms to control and mitigate the cascade. Hence, we develop the low complexity \emph{Cascading Failure Evolution -- Pseudo-inverse Based (CFE-PB) Algorithm} for identifying the evolution of a cascade that may be initiated by a failure of \emph{several} edges. The algorithm is based on the rank-1 update of the pseudo-inverse of the admittance matrix. We show that its complexity  is $O(|V|^3+|F_t^*||V|^2)$ ($|F_t^*|$ is the number of edges that eventually fail). Namely, if $t=|F_t^*|$ (one edge fails at each round), the complexity of the CFE-PB Algorithm is $O(\min\{|V|,t\})$ times lower than that of the CFE Algorithm.
The main advantage of the CFE-PB Algorithm is that it leverages the special structure of the pseudo-inverse to identify properties of the underlying graph and to recompute an instance of the pseudo-inverse from a previous instance.

Finally, we prove that the problems of finding the set of failures (of at most a given size) with the largest impact under different metrics, are NP-hard (or hard to approximate). For the problem of finding the set of initial failures of size $k$ that causes a cascade resulting with the minimum possible yield (\emph{minimum yield problem}), we introduce a very simple heuristic termed the Most Vulnerable Edges Selection -- Resistance distance Based (MVES-RB) Algorithm. We numerically show that solutions obtained by it lead to a much lower yield than the solutions obtain by selecting the initial edge failures randomly. Moreover, in some small graphs with a single edge failure, it obtains the optimal solution.

The main contributions of this paper are two fold. First, we provide new tools, based on matrix analysis for assessing the impact of a single edge failure. Using these tools, we (i) obtain upper bounds on the flow changes after a single failure, (ii) develop a fast algorithm for identifying the evolution of the cascade, and (iii) develop a heuristic algorithm for the minimum yield problem. Second, we analyze the cascade properties analytically and via simulations.

This paper is organized as follows. Section~\ref{sec:related-work} reviews related work. Section~\ref{sec:model} describes the power flow, cascade model, metrics, and the graphs used in the simulations. In Section~\ref{sec:Matrixprop}, we derive the properties of the admittance matrix of the grid. Section~\ref{sec:Failure Analysis} presents the effects of a single edge failure and Section~\ref{sec:properties}  provides the unique properties of the cascade. Section~\ref{sec: Algebraic View} introduces the CFE-PB Algorithm. Section~\ref{sec:hardness} discusses the hardness of the problems associated with the cascade and introduces the MVES-RB Algorithm. Section~\ref{sec:conclusion} provides concluding remarks and directions for future work. The proofs appear in the appendices.

%% file: Related.tex
\section{Related Work}
\label{sec:related-work}

Network vulnerability to attacks  has been thoroughly studied (e.g., \cite{Neumayer2011disasters,agarwal6403901,phillips1993network,Kleinberg2004NFD} and references therein). However, most previous computational work did not consider power grids and cascading failures. Recent work on cascades focused on
probabilistic failure propagation models (e.g., \cite{yeh, Chassin2005667,buldyrev2010catastrophic}, and references therein).
However, real cascades~\cite{Blackout,IndiaBlackout,SD_briefing} and simulation studies \cite{Bern2012ACM,SmartGridComm11rep} indicate that the cascade propagation is different than that predicted by such models.

In Sections~\ref{sec:Matrixprop} and \ref{sec: Algebraic View}, we use the admittance matrix of the grid to compute flows. This is tightly connected to the problem of \emph{solving Laplacian systems}. Solving these systems can be done with several techniques, including Gaussian elimination and LU factorization~\cite{golub2012matrix}. Recently,~\cite{christiano2011electrical} designed algorithms that use preconditioning, to provide highly precise approximate solutions to Laplacian systems in nearly linear time. However, this approach only provides approximate solutions and is not suitable for analytical studies of the effects of edge failures.

In Section~\ref{sec:Failure Analysis}, we obtain upper bounds on the flow changes after a single failure and study the robustness of graph classes based on \emph{resistance distance} and \emph{Kirchhoff's index}~\cite{klein1993resistance,bonchev1994molecular}. Recently, these notions have gained attention outside the Chemistry community. For instance, they were used in network science for detecting communities within a network, and more generally the strength of the connection between nodes in a network~\cite{newman2004finding,mcrae2006isolation}. Moreover, \cite{Hines2013Partition} recently used the resistance distance to partition power systems into zones.

The problem of \emph{identifying the set of failures with the largest impact} was studied in \cite{Bie10, Dan2,pinar_power,liu2013optimal}. In particular, \cite{Bie10} studies the $N-k$ problem which focuses on finding a small cardinality set of links whose removal disables the network from delivering a minimum amount of demand. A broader network interdiction problem in which all the components of the network are subject to failure was studied in~\cite{salmeron2004analysis}. A similar problem is studied in \cite{pinar_power} using the alternating-current (AC)  model. However, none of the previous works consider the cascading failures.
Moreover, while the optimal power flow problem  has been shown to be NP-hard~\cite{lavaei2012zero}, the complexity of the cascade-related problems was not studied yet.

Finally, for the simulations, we use \emph{graphs that can represent the topology of the power grid}. The structure of the power grids has been widely studied~\cite{watts1998collective,barabasi1999emergence,amaral2000classes,albert2004structural,crucitti2004topological,Chassin2005667,cotilla2012comparing}. In particular, Watts and Strogatz~\cite{watts1998collective} suggested the small-world graph as a good representative of the power grid,  based on the shortest paths between nodes and the clustering coefficient of the nodes. Barab\'{a}si and Albert~\cite{barabasi1999emergence,Chassin2005667} showed that scale-free graphs are better representatives based on the degree distribution. However, \cite{cotilla2012comparing} indicated that none of these models can represent U.S. Western interconnection properly.
Following these papers, we consider the Erd\H{o}s-R\'{e}nyi graph~\cite{erdHos1959random} in addition to these graphs.

%% file: Model.tex
\section{Models and Metrics}
\label{sec:model}

\vspace*{-0.2cm}
\subsection{DC Power Flow Model}
\label{ssec:flow-model}

We adopt the linearized (or DC) power flow model, which is widely used as an approximation for the more accurate non-linear AC power flow model~\cite{bergenvittal}.
In particular we follow ~\cite{Bern2012ACM,SmartGridComm11rep,Bie10,Dan2} and represent the power grid by an undirected graph $G=(V,E)$ where $V$ and $E$ are the set of nodes and edges corresponding to the buses and transmission lines, respectively.
$p_v$ is the active power \emph{supply} ($p_v>0$) or \emph{demand} ($p_v<0$) at node $v\in V$ (for a \emph{neutral node} $p_v=0$).
We assume \emph{pure reactive} lines, implying that each edge $\{u,v\} \in E$ is characterized by its \emph{reactance} $x_{uv}=x_{vu}>0$.

Given the power supply/demand vector $P\in \mathbb{R}^{|V|\times1}$ and the reactance values, a \emph{power flow} is a solution $(f, \theta)$ of:
\begin{eqnarray}
\label{eqn:flow1}&&\sum_{v \in N(u)}f_{uv} = p_u, \ \forall~ u \in V \\
\label{eqn:flow2}&&\theta_u - \theta_v - x_{uv}f_{uv} = 0, \ \forall~ \{u,v\} \in E
\end{eqnarray}
where $N(u)$ is the set of neighbors of node $u$, $f_{uv}$ is the power flow from node $u$ to node $v$, and $\theta_u$ is the phase angle of node $u$.
Eq.~(\ref{eqn:flow1}) guarantees (classical) flow conservation and (\ref{eqn:flow2}) captures the dependency of the flow on the reactance values and phase angles. Additionally, (\ref{eqn:flow2}) implies that $f_{uv}=-f_{vu}$.
\emph{Note that the edge capacities are not taken into account in determining the flows.}
When the total supply equals the total demand in each connected component of $G$, (\ref{eqn:flow1})-(\ref{eqn:flow2}) has a unique solution~\cite[lemma 1.1]{Bie10}.\footnote{The uniqueness is in the values of $f_{uv}$-s rather than $\theta_{u}$-s (shifting all $\theta_u$-s by equal amounts does not violate (\ref{eqn:flow2})).}
Eq.(\ref{eqn:flow1})-(\ref{eqn:flow2}) are equivalent to the following matrix equation:
\begin{equation}\label{eqn:flowmatrix}
A\Theta=P
\end{equation}
where $\Theta\in \mathbb{R}^{|V|\times 1}$ is the vector of phase angles and $A\in \mathbb{R}^{|V|\times |V|}$ is the \textit{admittance matrix} of the graph $G$, defined as follows:
\begin{equation*}\label{def: plus}
a_{uv}=
\begin{cases}
0&\text{if}~ u\neq v~\text{and}~\{u,v\}\notin E\\
-1/x_{uv}&\text{if}~u\neq v~\text{and}~\{u,v\}\in E\\
-\sum_{w\in N(u)} a_{uw}&\text{if}~u=v.
\end{cases}
\end{equation*}

If there are $k$ multiple edges between nodes $u$ and $v$, then $a_{uv}=-\sum_{i=1}^{k}1/x_{uv_i}$. Notice that when $x_{uv}=1 ~\forall\{u,v\}\in E$, the admittance matrix $A$ is the \emph{Laplacian matrix} of the graph~\cite{biggs1994algebraic}. Once $\Theta$ is computed, the power flows, $f_{uv}$, can be obtained from~(2).

Throughout this paper $\|.\|$  denotes the \textit{Euclidean norm} of the vector and the \emph{operator matrix norm}. For matrix $Q$, $q_{ij}$ denotes its $ij^{th}$ entry, $Q_i$ its $i^{th}$ row, and $Q^t$ its transpose.

\begin{algorithm}[t]
\caption{- Cascading Failure Evolution (CFE)}
\label{algorithm:cascade-2}
\vspace{-0.3cm}
\small
\begin{trivlist}
\item\textbf{Input:} A connected graph $G=(V,E)$ and an initial edge failures event $F_{0} \subseteq E$.
\end{trivlist}
\vspace*{-4mm}
\begin{algorithmic}[1]
\STATE $F^{*}_{0} \leftarrow F_{0}$ and $i \leftarrow 0$.
\WHILE{$F_{i} \neq \emptyset$}
\STATE  Adjust the total demand to equal the total supply within each connected component of $G = (V,E \setminus F^{*}_{i})$.
\STATE Compute the new flows $f_{e}(F^{*}_{i})$ ~ $\forall e \in E \setminus F^{*}_{i}$.
\STATE Find the set of new edge failures $F_{i+1}=\{e|f_{e}(F^{*}_{i})>c_e,~e\in E \setminus F^{*}_{i}\}$. $F^{*}_{i+1} \leftarrow F^{*}_{i} \cup F_{i+1}$ and $i \leftarrow i+1$.
\ENDWHILE
\RETURN $t = i-1$, $(F_{0}, \ldots, F_{t})$, and $f_{e}(F^{*}_{t})~ \forall e \in E\backslash F^{*}_{t}$.
\end{algorithmic}
\end{algorithm}

\subsection{Cascading Failure Model}
\label{ssec:cascade}

The Cascading Failure Evolution (CFE) Algorithm described here is a slightly simplified version of the cascade model used in~\cite{Dobson,SmartGridComm11rep,Bie10}.
We define $f_e=|f_{uv}|=|f_{vu}|$ and assume that an edge $e=\{u,v\}\in E$ has a predetermined power capacity $c_e=c_{uv}=c_{vu}$, which bounds its flow (that is, $f_e\leq c_e$).
The cascade proceeds in rounds. Denote by $F_i\subseteq E$  the set of edge failures in the $i^{th}$ round and by $F_{i}^{*}=F^{*}_{i-1} \cup F_{i}$ the set of edge failures until the end of the $i^{th}$ round ($i \geq 1$). We assume that before the initial failure event $F_0\subseteq E$, the power flows satisfy (\ref{eqn:flow1})-(\ref{eqn:flow2}), and $f_e\leq c_e~\forall e\in E$.
Upon a failure, some edges are removed from the graph, implying that it may become disconnected. Thus, within each component, the total demand is adjusted to be equal to the total supply.
For any set of failures $F \subseteq E$, we denote by $f_{e}(F)$ the flow along edges in $G'=(V,E \setminus F)$ after the load shedding.

\begin{figure*}[t]
\centering
\begin{subfigure}[h]{0.21\textwidth}
\centering
\raisebox{0cm}{\includegraphics[width=0.9\textwidth]{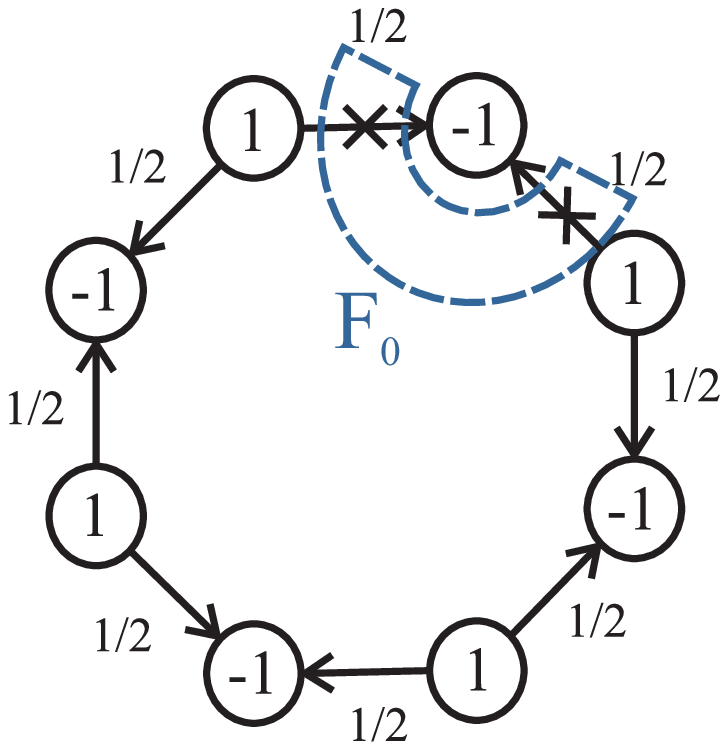}}
\caption{Initial flows and the failure event ($F_0$).}
\label{fig: Ex1Initial}
\end{subfigure}
\hspace{0.2cm}
\begin{subfigure}[h]{0.21\textwidth}
\centering
\includegraphics[width=0.9\textwidth]{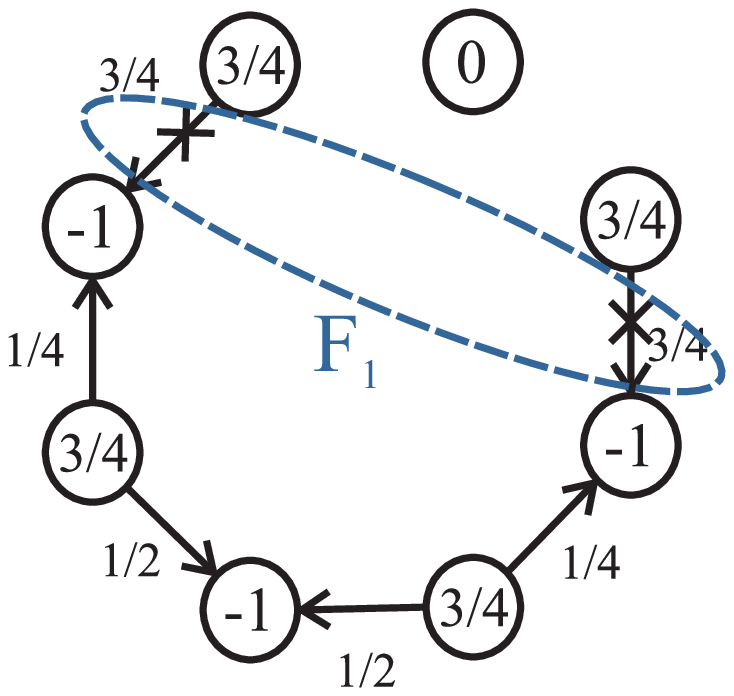}
\caption{Flows and failures due to overload ($F_1$).}
\vspace{-0.3cm}
\label{fig: EX1ClassicalFlow}
\end{subfigure}
\hspace{0.2cm}
\begin{subfigure}[h]{0.21\textwidth}
\centering
\hspace{-0cm}
\vspace{-0.85cm}
\raisebox{1.1cm}{\includegraphics[width=0.9\textwidth]{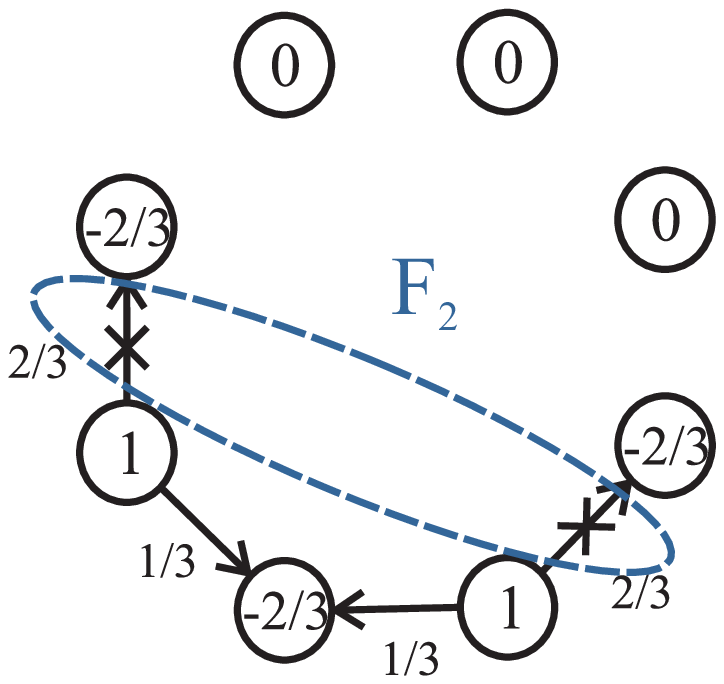}}
\vspace{-0.1cm}
\caption{Flows and failures ($F_2$).
}
\label{fig: EX1DCpowerFlowProp}
\end{subfigure}
\hspace{0.2cm}
\begin{subfigure}[h]{0.21\textwidth}
\centering
\vspace{-0.05cm}
\raisebox{0cm}{\includegraphics[width=0.9\textwidth]{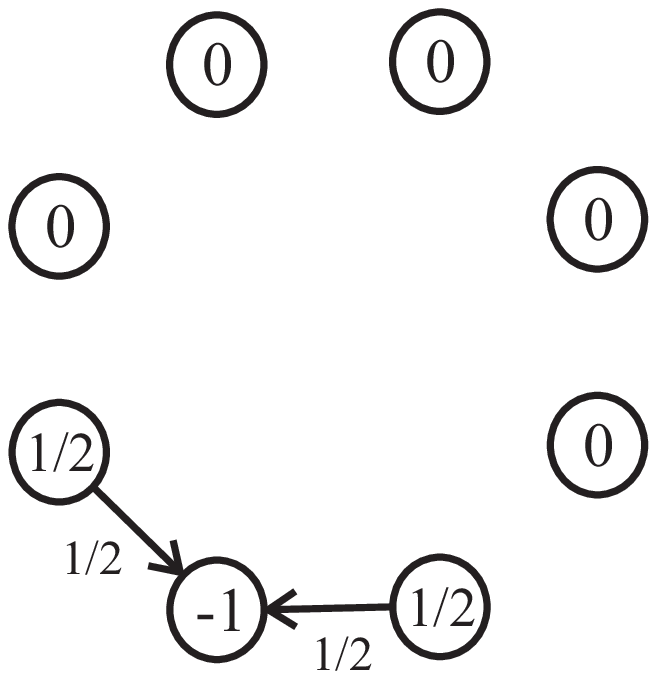}}
\vspace{0.05cm}
\caption{Stable state.\newline}
\vspace{-0.23cm}
\label{fig: EX1DCpowerFlow}
\end{subfigure}
\vspace{-0.3cm}
\caption{An example of a cascading failure  initiated by outages of the edges connecting a demand node to the network. The edge capacities and reactance values are $c_e=0.6$,  $x_e=1$. Numbers in nodes indicate power supply or demand ($p_v$), numbers on edges indicate flows ($f_e$), and arrows indicate flow direction.}
\label{fig:Example 1}
\vspace*{-0.3cm}
\end{figure*}

Following an initial failure event $F_{0}$, the new flows $f_{e}(F_{0}),\\ \forall e \in E\backslash F_0$ are computed (by (\ref{eqn:flow1})-(\ref{eqn:flow2})) (Line 4).
Then, the set of new edge failures $F_{1}$ is identified (Line 5).
Following~\cite{Dobson,SmartGridComm11rep,Bie10}, we use a deterministic outage rule and assume, for simplicity, that an edge $e$ fails once the flow exceeds its capacity: $f_{e}(F^{*}_{0}) > c_{e}$.\footnote{Note that~\cite{Dobson,SmartGridComm11rep,Bie10} maintain moving averages of the $f_e$ values to determine which edges fail.} Therefore, $F_{1} = \{e: f_{e}(F^{*}_{0}) > c_e, e \in E\backslash F_0^*\}$.

If the set $F_{1}$ of new edge failures is empty, then the cascade is terminated.
Otherwise, the process is repeated while replacing the initial event $F^{*}_{0} = F_{0}$ by the failure event $F^{*}_{1}$, and more generally replacing $F^{*}_{i}$ by $F^{*}_{i+1}$ at the $i^{th}$ round (Line 5).
The process continues until the system \textit{stabilizes}, namely until no edges are removed. Finally, we obtain the sequence $(F_{0}, F_{1}, \ldots, F_{t})$ of the sets of failures associated with the initial event $F_{0}$, and the power flows $f_e(F_t^*)$ at stabilization, where $t$ is the number of rounds until the network stabilizes. Since solving a system of linear equations with $n$ variables, requires $O(n^3)$ time~\cite{golub2012matrix}, the output can be obtained in $O(t|V|^3)$ time.

An example of a cascade can be seen in Fig.~\ref{fig:Example 1}. Initially, the flows are $f_e=0.5$ for all edges. The initial set of failures ($F_0$) disconnects a demand node from the graph.
Hence, intuitively, one may not expect a cascade.
However, this initial failure not only causes further failures but also causes failures in all edges except for two.
This example can be generalized to a graph with $2n$ nodes where with the same set of initial failures, all the edges fail except for two.

For simplicity, when the initial failure event contains a single edge, $F_0=\{e'\}$, we denote the flows after the failure by $f_e' \equiv f_e(\{e'\})$ and the flow changes by $\Delta f_{e}=f_e'-f_e~\forall e\in E\backslash\{e'\}$.

\subsection{Metrics}
\label{ssec:metrics}

We define the metrics for evaluating the grid vulnerability (some of which were defined in~\cite{SmartGridComm11rep}).
To study the effects of a \emph{single edge ($e'$) failure after one round}, we define the ratio between the change of flow on an edge, $e$, and its original value or the flow value on the failed edge, $e'$:

\noindent \textbf{Edge flow change ratio}: $S_{e,e'}=|\Delta f_e/f_{e}|$.\\
\noindent \textbf{Mutual edge flow change ratio}: $M_{e,e'}=|\Delta f_{e}/f_{e'}|$.\\
\vspace*{-0.3cm}

Below, we define metrics related to the evaluation of the \emph{cascade severity} for a given instance $G$, an initial failure event $F_{0} \subseteq E$, and an integer $k \geq 1$. An instance is composed of a connected graph $G$, supply/demand vector $P$, capacities and reactance values $c_e$,  $x_e$ $\forall e \in E$.
For brevity, an instance is represented by $G$.

\noindent\textbf{Yield} (the ratio between the demand supplied at stabilization and the original demand):
$Y(G,F_{0})$, $Y(G,k) = \min_{F_{0} \subseteq E, |F_{0}| \leq k} Y(G,F_{0})$.

\noindent\textbf{Number of edge failures}: $|F^{*}_{+}(G,F_{0})|$, \\$|F^{*}_{+}(G,k)|=\max_{F_{0} \subseteq E, |F_{0}| \leq k} |F^{*}_{+}(G,F_{0})|$.

\noindent\textbf{Number of rounds until stabilization}: $L(G,F_{0})$, \\$L(G,k)=\max_{F_{0} \subseteq E, |F_{0}| \leq k} L(G,F_{0})$.

 For the following metric, we define: (i) $d(e,e')$ as the distance (in hop count) between edges $e$ and $e'$ in $G$, and (ii) for any $i$, $ d(F_{i-1},F_{i})= \min_{e \in F_{i-1},e' \in F_{i}} d(e,e')$.

\noindent\textbf{Distance between failures}:  $D(G,F_{0})=\min_{i, 1
\leq i \leq t}\\ d(F_{i-1},F_{i})$, $D(G,k) = \max_{F_{0} \subseteq E,
|F_{0}| \leq k} D(G,F_{0})$.\\
\vspace*{-0.4cm}
\subsection{Graphs Used in Simulations}\label{subsec:graph}
The simulation results are presented for the graphs described below. All graphs have 1,374 nodes to correspond the subgraph of the Western interconnection. The parameters are as indicted below, unless otherwise mentioned.

\noindent \textbf{Western interconnection}: 1708-edge connected subgraph of the U.S. Western interconnection. The data is from
the Platts Geographic Information System~\cite{GIS}.\\
\noindent \textbf{Erd\H{o}s-R\'{e}nyi graph}~\cite{erdHos1959random}: A random graph where each edge appears with probability $p=0.01$.\\
\noindent \textbf{Watts and Strogatz graph}~\cite{watts1998collective}: A small-world random graph where each node connects to $k=4$ other nodes and the probability of rewiring is $p=0.1$.\\
\noindent \textbf{Bar\'{a}basi and Albert graph}~\cite{barabasi1999emergence}: A scale-free random graph where each new node connects to $k=3$ other nodes at each step following the preferential attachment mechanism.

%% file: MatrixProperties.tex
\section{Admittance Matrix Properties}\label{sec:Matrixprop}

 In this section,  we use the \textit{Moore-Penrose Pseudo-inverse} of the admittance matrix~\cite{albert1972regression} in order to obtain results that are used throughout the rest of the paper. Specifically they are used in Section~\ref{sec:Failure Analysis} to study the impact of a single edge failure on the flows on other edges and in Section~\ref{sec: Algebraic View} to introduce an efficient algorithm to identify the evolution of the cascade.
We prove several properties of the Pseudo-inverse of the admittance matrix $A$, denoted by $A^+$.\footnote{$A^+=\lim_{\delta \to 0} A^t (A A^t+\delta^2 I)^{-1}$~\cite{albert1972regression}. For more information regarding the definition, see Appendix.} $A^+$ always exists regardless of the structure of the graph $G$. Some proofs and results that are used in the proofs appear in Appendix~\ref{app:prelem}.

\begin{figure*}[t]
\vspace*{-0.1cm}
\centering
\includegraphics[scale=0.33]{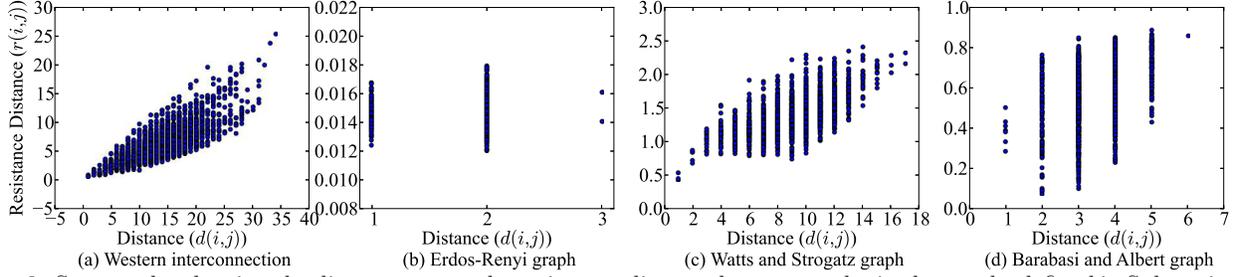}
\vspace*{-0.8cm}
\caption{Scatter plot showing the  distance versus the resistance distance between nodes in the graphs defined in Subsection~\ref{subsec:graph}.}
\vspace*{-0.3cm}
\label{figure:distance vs res distance}
\end{figure*}

Observation~\ref{lem: pseudo inv matrix sol} shows that the power flow equations can be solved by using $A^+$.
\vspace*{-0.2cm}
\begin{observation}\label{lem: pseudo inv matrix sol}
If (\ref{eqn:flowmatrix}) has a feasible solution, $\hat{\Theta}=A^+P$ is a solution for (\ref{eqn:flowmatrix}).\footnote{Recall from Section~\ref{sec:model} that (\ref{eqn:flow1})-(\ref{eqn:flow2}) have a unique solution with respect to power flows but not in respect to phase angles. Therefore, the solution to (\ref{eqn:flowmatrix}) may not be unique.}
\end{observation}
\vspace*{-0.2cm}
\vspace*{-0.2cm}
\begin{proof}
According to Theorem \ref{th: pseudo inverse}, $\hat{\Theta}=A^+P$ minimizes $\|P-A\Theta\|$. On the other hand, since (\ref{eqn:flowmatrix}) has a solution, $\|P-A\hat{\Theta}\| =\min_{\Theta} \|P-A\Theta\|=0$. Thus, $\hat{\Theta}=A^+P$ is a solution for (\ref{eqn:flowmatrix}).
\end{proof}
\vspace*{-0.2cm}
Jointly verifying whether an edge is a cut-edge and finding the connected components of the graph takes $O(|E|)$ (using Depth First Search~\cite{cormen2001introduction}). The following two Lemmas show that by using the precomputed pseudo-inverse of the admittance matrix, these operations can be done in $O(1)$ and $O(|V|)$, respectively. The algorithm in Section~\ref{sec:algorithm} uses the results to check if the pseudo-inverse should be recomputed. Moreover, Lemma~\ref{lem:cutedge find} is crucial for the proof of the Theorem~\ref{col: pseudo inv}, below.
\vspace*{-0.2cm}
\begin{lemma}[Bapat~\cite{bapat2010graphs}]\label{lem:cutedge find}
Given $G=(V,E)$ and $A^+$, all the \textit{cut-edges} of the graph $G$ can be found in $O(|E|)$ time. Specifically, an edge $\{i,j\}\in E$ is a cut-edge if, and only if, $a_{ij}^{-1}-2a_{ij}^++a_{ii}^++a_{jj}^+=0$.
\end{lemma}
\vspace*{-0.2cm}
\vspace*{-0.2cm}
\begin{lemma}\label{lem:connected comp find}
Given $G=(V,E)$, $A^+$, and the cut-edge $\{i,j\}$, the connected components of $G\backslash \{i,j\}$ can be identified in $O(|V|)$.
\end{lemma}
\vspace*{-0.2cm}
In the following, we denote by $A'$ the admittance matrix of the graph $G'=(V,E\backslash\{i,j\})$ and by $P'$ the power vector after removing an arbitrary edge $e'=\{i,j\}$ from the graph $G$ and conducting the corresponding load shedding.

Lemma~\ref{lem:cutedge pseudo} shows that after the removal of a cut-edge, $A^+$ can be used to solve~(\ref{eqn:flowmatrix}) and $A'^+$ is not required.
\vspace*{-0.2cm}
\begin{lemma}\label{lem:cutedge pseudo}
Given graph $G=(V,E)$, $A^+$, and a cut-edge $\{i,j\}$, then $\hat{\Theta}=A^+P'$ is a solution of~(\ref{eqn:flowmatrix}) in $G'$.
\end{lemma}
\vspace*{-0.2cm}
The following theorem gives an analytical rank-1 update of the pseudo-inverse of the admittance matrix. Using Theorem~\ref{col: pseudo inv} and Corollary~\ref{col: pseudo inv change}, in Section~\ref{sec:Failure Analysis} we provide upper bounds on the mutual edge flow change ratios ($M_{e,e'}$). We note that a similar result to Theorem~\ref{col: pseudo inv} was independently proved in a very recent technical report~\cite{ranjan2013incremental}.
\vspace*{-0.2cm}
\begin{theorem} \label{col: pseudo inv}
Given graph $G=(V,E)$, the admittance matrix $A$, and $A^+$, if $\{i,j\}$ is not a cut-edge, then,
\begin{equation*}
A'^+=(A+a_{ij}XX^t)^+ = A^+ - \frac{1}{a_{ij}^{-1}+X^tA^+X} A^+XX^tA^+
\end{equation*}
in which $X$ is an $n\times1$ vector with $1$ in $i^{th}$ entry, $-1$ in $j^{th}$ entry, and 0 elsewhere.
\end{theorem}
\vspace*{-0.2cm}
For the following, recall from Section~\ref{sec:model} that $A^+=[a^+_{rs}]$.
\vspace*{-0.2cm}
\begin{corollary}\label{col: pseudo inv change}
\begin{equation*}
f'_{rs}=f_{rs}-\frac{a_{rs}}{a_{ij}}\frac{(a^{+}_{ri}-a^{+}_{rj})-(a^{+}_{si}-a^{+}_{sj})}{a_{ij}^{-1}-2(a^{+})_{ij}+(a^{+})_{ii}+(a^{+})_{jj}}f_{ij}.
\end{equation*}
\end{corollary}
\vspace*{-0.2cm}
Finally, Lemma~\ref{lem:update pseudo inverse}, gives the complexity of the rank-1 update provided in Theorem~\ref{col: pseudo inv}. This is used in the computation of the running time of the algorithm in Section~\ref{sec: Algebraic View}.
\vspace*{-0.2cm}
\begin{lemma}\label{lem:update pseudo inverse}
Given graph $G=(V,E)$, $A^+$, and an edge $\{i,j\}$, which is not a cut-edge of the graph, $A'^+$ can be computed from $A^+$ in $O(|V|^2)$.
\end{lemma}
\vspace*{-0.2cm}

 We now define the notion \emph{resistance distance}~\cite{klein1993resistance}. In resistive circuits, the resistance distance between two nodes is the equivalent resistance between them. It is known that the resistance distance, is actually a measure of distance between nodes of the graph~\cite{bapat2010graphs}. For any network, this notion can be defined by using the pseudo-inverse of the Laplacian matrix of the network. Specifically, it can be defined in power grid networks by using the pseudo-inverse of the admittance matrix, $A^+$.
\vspace*{-0.2cm}
\begin{definition}\label{def:resistance distance}
Given $G=(V,E)$, $A$, and $A^+$, the \emph{resistance distance} \emph{between two nodes} $i,j\in V$ is $r(i,j):= a_{ii}^++a_{jj}^+-2a_{ij}^+$. Accordingly, the resistance distance \emph{between two edges} $e=\{i,j\},e'=\{p,q\}$ is $r(e,e')=\min\{r(i,p),r(i,q),\\r(j,p),r(j,q)\}$.
\end{definition}
\vspace*{-0.2cm}
When all the edges have the same reactance, $x_{ij}=1~\forall \{i,j\}\in E$, the resistance distance between two nodes is a measure of their connectivity. Smaller resistance distance between nodes $i$ and $j$ indicates that they are better connected. Fig.~\ref{figure:distance vs res distance} shows the relation between the distance and the resistance distance between nodes in the graphs defined in Subsection~\ref{subsec:graph} (all the edges have the reactance equal to 1). As can be seen, there is no direct relation between these two measures in Erd\H{o}s-R\'{e}nyi and Bar\'{a}basi-Albert graphs. However, in the Western interconnection and Watts-Strogatz graph the resistance distance increases with the distance.

 In Chemistry, the sum over the resistance distances between all pairs of nodes in the graph $G$ is referred to as the \emph{Kirchhoff index}~\cite{bonchev1994molecular} of $G$ and denoted by $Kf(G)$. We use this notion in Subsection~\ref{subsubsec: Robust Structures} to study the robustness of different graph classes to single edge failures.
 \vspace*{-0.2cm}
 \begin{definition}\label{def:Kf}
 Given $G=(V,E)$ and $A$, the \emph{Kirchhoff index} of $G$ is $Kf(G)=\frac{1}{2}\sum_{i,j\in V} r(i,j)$.
 \end{definition}
 \vspace*{-0.2cm} 

%% file: FailureAnalysis.tex
\section{Effects of a Single Edge Failure}\label{sec:Failure Analysis}
In this section we provide upper bounds on the flow changes after a single edge failure and study the robustness of different graph classes.
\begin{figure*}[t]
\centering
\includegraphics[scale=0.28]{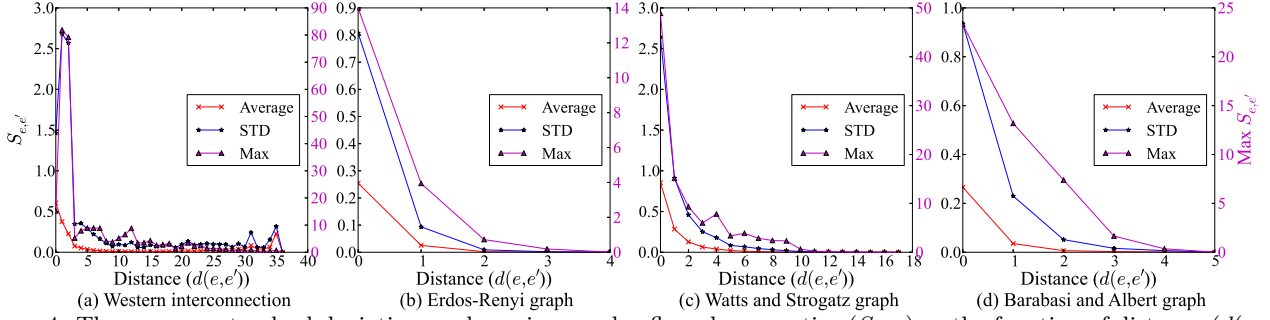}
\vspace*{-0.4cm}
\caption{The average, standard deviation, and maximum edge flow change ratios ($S_{e,e'}$) as the function of distance ($d(e,e')$) from the failure. The right $y$-axis shows the values for the maximum edge flow change ratios ($\max S_{e,e'}$). The data points are obtained for 40 different random choices of an initial edge failure.}
\vspace*{-0.5cm}
\label{figure:Delta Flow over in}
\end{figure*}
For simplicity, in this section, we assume that $x_e=1~\forall e\in E$, unless otherwise indicated.  As mentioned in Section~\ref{sec:model}, in this case the \emph{admittance matrix} of the graph, $A$, is equivalent to the \emph{Laplacian matrix} of the graph. However, all the results can be easily generalized.
\subsection{Flow Changes}
\subsubsection{Edge Flow Change Ratio}
In order to provide insight into the effects of a single edge failure, we first present simulation results. The simulations have been done in Python using NetworkX library. Fig.~\ref{figure:Delta Flow over in} shows the edge flow change ratios ($S_{e,e'}$) as the function of distance ($d(e,e')$) from the failure for over 40 different random choices of an initial edge failure, $e'$. The power supply/demand in the Western interconnection is based on the actual data. In other graphs, the power supply/demand at nodes are i.i.d. Normal random variables with a slack node to equalize the supply and demand. Notice that if the initial flow in an edge is close to zero, the edge flow change ratio on that edge can be very large. Thus, to focus on the impact of an edge failure on the edges with reasonable initial flows, we do not illustrate the edge flow change ratios for the edges with flow below 1\% of the average flow. Yet, we observe that such edges that experience a flow increase after a single edge failure, are within any arbitrary distance from the initial edge failure.

Fig.~\ref{figure:Delta Flow over in} shows that after a single edge failure, there might be a very large increase in flows (edge flow change ratios up to 80, 14, 50, and 24  in Fig.~\ref{figure:Delta Flow over in}-(a), (b), (c), and (d), respectively) and sometimes far from the initial edge failure (edge flow change ratio around 10 for edges 11- and 4-hops away from the initial failure in Fig.~\ref{figure:Delta Flow over in}-(a) and (c), respectively). Moreover, as we observed in all of the four graphs, there are edges with positive flow increase from zero, far from the initial edge failure. These observations motivate us to prove similar results analytically (see Observation~\ref{lemma:grand-flow-increase} in this section and Observation~\ref{lemma:grande-taille} and \ref{lemma:grande-distance-repetition} in Section~\ref{sec:properties}).

Finally, we show that by choosing the parameters in a specific way, the edge flow change ratio can be arbitrarily large.
\vspace*{-0.2cm}
\begin{observation}
\label{lemma:grand-flow-increase}
For any $x_{e_1},x_{e_2}\in \mathbb{R}$, there exists a graph $G=(V,E)$ and two edges $e_{1},e_{2} \in E$ such that \\$S_{e_2,e_1} = x_{e_2}/x_{e_1}$.
\end{observation}
\vspace*{-0.2cm}
\subsubsection{Mutual Edge Flow Change Ratio}
We use the notion of \emph{resistance distance} to find upper bounds on the mutual edge flow change ratios ($M_{e,e'}$).
The following Lemma provides a formula for computing the flow changes after a single edge failure based on the resistance distances. It is \emph{independent} of the power supply/demand distribution.
\vspace*{-0.2cm}
\begin{lemma}\label{lem:flow changes}
Given $G=(V,E)$, $A$, and $A^+$, the flow change and the mutual edge flow change ratio for an edge $e=\{i,j\}\in E$ after a failure in a non-cut-edge $e'=\{p,q\}\in E$ are,
\begin{eqnarray*}
\Delta f_{ij}&=&\frac{1}{2}\frac{-r(i,p)+r(i,q)+r(j,p)-r(j,q)}{1-r(p,q)}f_{pq},\\
M_{e,e'}&=& \frac{1}{2}\frac{-r(i,p)+r(i,q)+r(j,p)-r(j,q)}{1-r(p,q)}.
\end{eqnarray*}
\end{lemma}
\vspace*{-0.2cm}
\begin{proof}
It is an immediate result of Corollary~\ref{col: pseudo inv change}.
\end{proof}
\vspace*{-0.2cm}
\begin{figure}[t]
\vspace*{-0.17cm}
\centering
\includegraphics[scale=0.38]{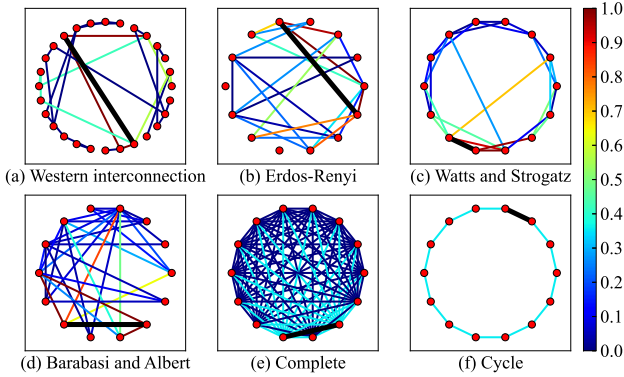}
\vspace*{-0.6cm}
\caption{The mutual edge flow change ratios ($M_{e,e'}$) after an edge failure (represented by black wide line) in different graph classes. All the graphs have 12 nodes, except for (a) which has 28 nodes. In (b) $p=0.1$.}
\vspace*{0.1cm}
\label{figure:Flow Changes Comparison}
\end{figure}
Fig.~\ref{figure:Flow Changes Comparison} illustrates  the mutual edge flow change ratios after an edge failure. Recall that $M_{e,e'}$ describes the distribution of the flow that passed through $e'$ on the other edges. These values are differently distributed for different graph classes. In the next subsection, we study in detail the relation between the mutual edge flow change ratios and the graph structure.

The following Corollary gives an upper bound on the flow changes after a failure in a non-cut-edge $\{p,q\}\in E$ by using the triangle inequality for resistance distance and Lemma~\ref{lem:flow changes}.
\vspace*{-0.2cm}
\begin{corollary}\label{col:upper bound1}
Given $G=(V,E)$, $A$, and $A^+$, the flow changes in any edge $e=\{i,j\}\in E$ after a failure in a non-cut-edge $e'=\{p,q\}\in E$ can be bounded by,
\begin{equation*}
|\Delta f_{ij}|\leq \frac{r(p,q)}{1-r(p,q)}|f_{pq}|,~M_{e,e'}\leq \frac{r(p,q)}{1-r(p,q)}.
\end{equation*}
\end{corollary}
\vspace*{-0.2cm}

With the very same idea, the following corollary gives an upper bound on the flow changes in a specific edge $\{i,j\}\in E$ after a failure in the non-cut-edge $\{p,q\}\in E$.
\vspace*{-0.4cm}
\begin{corollary}\label{col:upper bound2}
Given $G=(V,E)$, $A$, and $A^+$, the flow changes in an edge $e=\{i,j\}\in E$ after a failure in a non-cut-edge $e'=\{p,q\}\in E$ and the mutual edge flow change ratio $M(e,e')$ can be bounded by,
\begin{equation*}
|\Delta f_{ij}|\leq \frac{r(e,e')}{1-r(p,q)}|f_{pq}|,~M_{e,e'}\leq\frac{r(e,e')}{1-r(p,q)}.
\end{equation*}
\end{corollary}
\vspace*{-0.2cm}
Corollary~\ref{col:upper bound2}  directly connects the resistance distance between two edges ($r(e,e')$) to their mutual edge flow change ratio ($M_{e,e'}$). It shows that the resistance distance, in contrast to the distance, can be used for assessing the influence of an edge failure on other edges.

\begin{figure*}[t]
\vspace*{-0.2cm}
\centering
\includegraphics[scale=0.28]{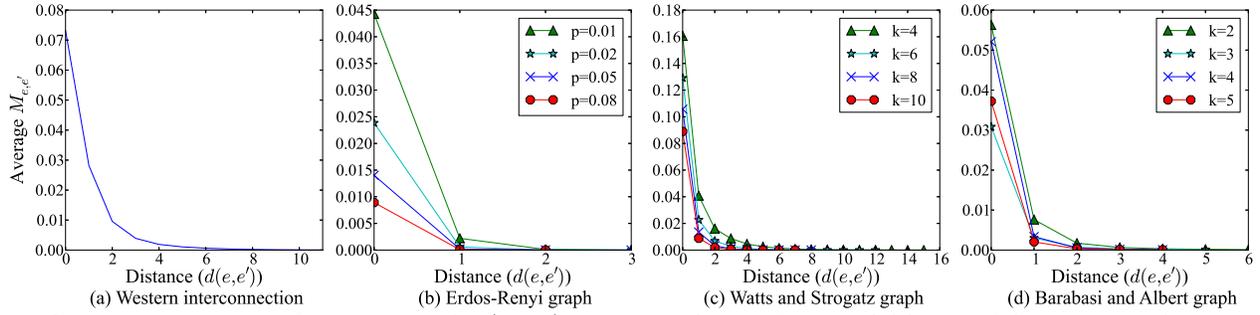}
\vspace*{-0.4cm}
\caption{The average mutual flow change ratios ($M_{e,e'}$) versus the distance from the initial edge failure. Each point represents the average of 40 different initial single edge failure events.}
\vspace{-0.2cm}
\label{figure:Flow Changes vs distance Comparison}
\end{figure*}

\begin{figure*}[t]
\vspace{-0.1cm}
\centering
\includegraphics[scale=0.28]{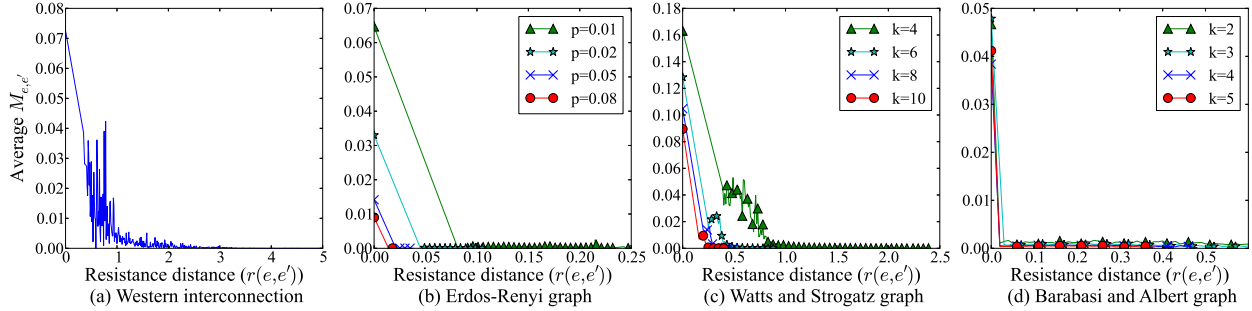}
\vspace*{-0.4cm}
\caption{The average mutual flow change ratios ($M_{e,e'}$) versus the resistance distance from the initial edge failure. Each point represents the average of 40 different initial single edge failure events. For clarity, the markers appear for every 5 data points.}
\vspace*{-0.4cm}
\label{figure:Flow Changes vs res distance Comparison}
\end{figure*}

We present simulations to show the relations between the mutual flow change ratios and the two distance measures. Figs.~\ref{figure:Flow Changes vs distance Comparison} and \ref{figure:Flow Changes vs res distance Comparison} show the mutual edge flow change ratio ($M_{e,e'}$) as the function of distance ($d(e,e')$) and resistance distance ($r(e,e')$) from the failure, respectively. The figures show that increasing number of edges (increasing $p$ in Erd\H{o}s-R\'{e}nyi graph and increasing $k$ in Watts and Strogatz, and Bar\'{a}basi and Albert graphs) affects the $M_{e,e'}$-$r(e,e')$ relation more than the $M_{e,e'}$-$d(e,e')$ relation. This suggests that the resistance distance better captures the information hidden in the structure of a graph.  Both figures show a monotone relation between the mutual edge flow change ratios and the distances/resistance distances. However, this monotonicity is smoother in the case of the distance.

Moreover, Fig.~\ref{figure:Flow Changes vs distance Comparison}, unlike Fig.~\ref{figure:Delta Flow over in}, shows that after a single edge failure, the mutual edge flow change ratios decrease as the distance from the initial failure increases. Thus, it suggests that probabilistic tools may be used to model the mutual edge flow change ratios ($M_{e,e'}$) better than the edge flow change ratios ($S_{e,e'}$).

\subsection{Graph Robustness}\label{subsec:structure}
We now use the upper bounds provided in Corollaries~\ref{col:upper bound1} and \ref{col:upper bound2} to study the robustness of some well-known graph classes to single edge failures. We use the average mutual edge flow change ratio, $M_{e,e'}$, as the measure of the robustness. The small value of $M_{e,e'}$ indicates that the flow changes in edges after a single edge failure is small compared to the original flow on the failed edge. In other words, the network is able to distribute additional load after a single edge failure uniformly between other edges.

We show that (i) graphs with more edges are more robust to single edge failures and (ii) the Kirchhoff index can be used as a measure for the robustness of different graph classes.

\subsubsection{Robustness Based on Number of Edges}\label{subsec:Robust Number of edges}
Using Corollary~\ref{col:upper bound1}, it can be seen that a failure in an edge with small resistance distance between its two end nodes leads to a small upper bound on the mutual edge flow change ratios, $M_{e,e'}$, on the other edges. Thus, the average $r(i,j)$ for $\{i,j\}\in E$ is relatively a good measure of the average mutual edge flow change ratio. The following Observation shows that graphs with more edges have smaller average $r(i,j)$ for $\{i,j\}\in E$, and therefore, smaller average mutual edge flow change ratio.
\vspace*{-0.2cm}
\begin{observation}\label{lem:av resisdis1}
Given $G=(V,E)$, the average $r(i,j)$ for $\{i,j\}\in E$ is $\frac{|V|-1}{|E|}$.
\end{observation}
\vspace*{-0.2cm}
Observation~\ref{lem:av resisdis1} implies that for a fixed number of nodes, the average resistance distance gets smaller as the number of edges increases. Therefore, graphs with more edges are more robust against a single edge failure.

\subsubsection{Robustness Based on the Graph Class} \label{subsubsec: Robust Structures}
Another way of computing the average mutual edge flow change ratio is to use Corollary~\ref{col:upper bound2} which implies that graphs with low average resistance distance over all pairs of nodes have the small average mutual edge flow change ratios. On the other hand, recall from Definition~\ref{def:Kf} that the average resistance distance over all pair of nodes is equal to Kirchhoff index of the graph divided by the number of edges. Hence, table~\ref{tab:tab Kf} summarizes the Kirchhoff indices and corresponding average mutual edge flow change ratios for some well-known graph classes.

\begin{table}[t]
\begin{center}
\vspace*{-0.2cm}
\caption{The Kirchhoff indices and the average mutual edge flow change ratios ($M_{e,e'}$) for some well-known graphs. The values that were previously known~\cite{lukovits1999resistance} are highlighted by grey cells.}
\vspace*{-0.2cm}
\scriptsize
\begin{tabular}{|m{2cm}|C{2.3cm}|C{2.8cm}|}
\hline
\textbf{Graph Class} & \textbf{Kirchhoff index} & \textbf{Average mutual edge flow change ratio ($M_{e,e'}$)}  \\
\hline
Complete graph & $n-1$ \cellcolor[gray]{0.8}& $O(\frac{1}{n})$  \\
\hline
Complete \mbox{bipartite} graph  & $4n-3$ \cellcolor[gray]{0.8}& $O(\frac{1}{n})$ \\
\hline
Complete \mbox{tripartite} graph  & $\frac{1}{2}(9n-5)$ \cellcolor[gray]{0.8}& $O(\frac{1}{n^2})$ \\
\hline
Cycle graph & $\frac{1}{12}(n-1)n(n+1)$ \cellcolor[gray]{0.8}& $O(n^2)$ \\
\hline
Cocktail party graph & $\frac{2n^2-2n+1}{n-1}$ \cellcolor[gray]{0.8}& $O(\frac{1}{n})$ \\
\hline
Erd\H{o}s-R\'{e}nyi graph & $\Theta(\frac{n}{p})$& $O(\frac{1}{np^2})$\\
\hline
\end{tabular}
\vspace*{-0.1cm}
\label{tab:tab Kf}
\end{center}
\end{table}
To complete the table, in the following lemma we compute the Kirchhoff index of the Erd\H{o}s-R\'{e}nyi graph as a function of $p$.
\vspace*{-0.2cm}
\begin{lemma}\label{lem: Erdos Renyi Kf}
For an Erd\H{o}s-R\'{e}nyi random graph, $G(n,p)$, $Kf(G)$ is of $\Theta(\frac{n}{p})$, and therefore the average resistance distance between all pairs of nodes is of $\Theta(\frac{1}{np^2})$.
\end{lemma}
\vspace*{-0.2cm}
 This Lemma shows that the average resistance distance between all pairs of nodes of an Erd\H{o}s-R\'{e}nyi graph is related to $1/p^2$. Since as $p$ grows, the average number of edges in a Erd\H{o}s-R\'{e}nyi graph increases, this Lemma also suggests that graphs with more edges are more robust to a single edge failure. Thus, the results in this subsection are aligned with the result in Subsection~\ref{subsec:Robust Number of edges} indicating that graphs with more edges are more robust to a single edge failure.

%% file: Properties.tex
\section{Properties of the Cascade}
\label{sec:properties}
As shown in the previous section, due to the special structure of the power flow equations, even studying the impact of a single edge failure is not straightforward. In this section, we focus on the cascade that can be caused by a single or multiple failures.
Using the metrics from Section~\ref{ssec:metrics}, we show unique properties of cascade.

\subsection{Non-monotone Effect of Failures}
\label{ssec:non-monotonicity}

We show that a single edge failure event $F_{0}=\{e\}$ may have a larger effect in terms of number of rounds, number of edge failures, and yield than any failure event $F$ that is a superset of $F_{0}$.
\vspace{-0.2cm}
\begin{observation}
\label{lemma:subset-1}
There exists a graph $G=(V,E)$, an initial failure $F_{0} = \{e\}$, $e \in E$, and $F'_0 \supset F_{0}$, such that $(V,E \setminus F'_0)$ is a connected graph, $Y(G,F_{0}) = 0$, and $Y(G,F'_0) = 1$.
\end{observation}
\vspace{-0.2cm}
\noindent This Observation implies that the identification of an initial failure event with the largest impact, is hard (see Section~\ref{sec:hardness}). In general, we cannot avoid considering a set of failures only because it is a subset of another set. We note that it is shown in~\cite[lemma 4.3]{SmartGridComm11rep} that an initial failure event $F_0$
may result in a lower yield than a failure event $F\supset F_0$. However, in the special case used in~\cite{SmartGridComm11rep}, $G\backslash F$ is disconnected. In Observation~\ref{lemma:subset-1}, we show that there exists a graph $G$ such that even when $G\backslash F$ is connected, a single edge failure event $F_0$ causes more damage than $F\supset F_0$.

\subsection{Unbounded Metric Values}
\label{ssec:unbounded}

By using simple instances, we show that the effect of a single edge failure may be arbitrarily severe (Table~\ref{tab:tab-1} summarizes the results).
First, we show that a single edge failure event may cause a cascading failure in which the number of cascade rounds is at the order of the number of edges, the yield is $0$, and all edges fail.
\begin{table}[t]
\begin{center}
\caption{Worst case values of the metrics for cascades caused by a single edge failure.}
\vspace*{-0.2cm}
\scriptsize
\begin{tabular}{|m{3.6cm}|c|c|c|}
\hline
\multicolumn{2}{|c|}{\textbf{Metric}} & \multicolumn{2}{c|}{\textbf{Worst case}} \\
\hline
Edge flow change ratio & $S_{e,e'}$ & $x_{e_1}/x_{e_2}$ & Obs.~\ref{lemma:grand-flow-increase} \\
\hline
Number of edge failures & $|F^{*}_{+}(G,1)|$ & $|E|$ & Obs.~\ref{lemma:grande-taille} \\
\hline
Number of rounds & $L(G,1)$ & $|E|-1$ & Obs.~\ref{lemma:grande-taille} \\
\hline
Yield & $Y(G,1)$ & $0$ & Obs.~\ref{lemma:grande-taille} \\
\hline
Distance between failures & $D(G,1)$ & $O(|E|)$ & Obs.~\ref{lemma:grande-distance-repetition} \\
\hline
\end{tabular}
\vspace*{-0.3cm}
\label{tab:tab-1}
\end{center}
\end{table}
\vspace*{-0.2cm}
\begin{observation}
\label{lemma:grande-taille}
For any integer $m$, there exists a graph $G=(V,E)$ with $|E|\geq m$, such that $L(G,1) = |E|-1$, $|F^{*}_{+}(G,1)| = |E|$, and $Y(G,1)=0$.
\end{observation}
\vspace*{-0.2cm}

Then, we show that cascading failures may happen within arbitrarily long distance of each other and may last arbitrarily long time. This corresponds to the simulation results in Fig.~\ref{figure:Delta Flow over in} that shows a single edge failure can have a very significant impact on the flows on far edges.  In~\cite[lemma 4.2]{SmartGridComm11rep} it was shown
that cascading failures may happen within arbitrarily long distance
of each other, and in~\cite[lemma 4.7]{SmartGridComm11rep} it was
shown that they can last arbitrarily long time. Yet,
we show that these two
events can happen  simultaneously.
\vspace*{-0.2cm}
\begin{observation}
\label{lemma:grande-distance-repetition}
For any $l, d \geq 1$, there exists a graph $G=(V,E)$ such that $L(G,1) \geq l$ and for any $i$, $1 \leq i \leq l$, $d_{i} \geq d$.
As a result $D(G,1) \geq d$.
\end{observation}
\vspace*{-0.2cm}
\subsection{Effects of Small Parameter Changes}
\label{ssec:small-changes}
We analyze the effect of very small changes in capacity, $c_e$, or reactance, $x_e$, of a single edge.
We show that a failure event that has negligible effects on the original instance can have a major impact for slightly modified instances. Let $\varepsilon > 0$ and define graphs $G^{c}_{-}$ and $G^{x}_{-}$ as the replications of the graph $G=(V,E)$ with a small difference in a parameter value of an edge $e \in E$.
In $G^{c}_{-}$, $c^{-}_e=c_e-\varepsilon$; and in $G^{x}_{-}$, $x^{-}_e=x_e-\varepsilon$.
We consider the consequences of a cascade caused by a single edge failure event $F_{0}$ ($|F_{0}|=1$), for $G$, $G^{c}_{-}$, and $G^{x}_{-}$.
\vspace*{-0.2cm}
\begin{observation}
\label{lemma:small-change-1}
For any $\varepsilon > 0$ and any integer $m$, there exists a graph $G=(V,E)$ with $|E| \geq m$, an edge $e\in E$, and an initial failure $F_{0} \subseteq E$ with $|F_{0}|=1$, such that: \\
$L(G,F_{0}) = 0$, $|F^{*}_{+}(G,F_{0})| = |F_{0}| = 1$, $Y(G,F_{0}) = 1$; \text{but}
{\emph{(i)}}~$L(G^{c}_{-},F_{0}) = |F^{*}_{+}(G^{c}_{-},F_{0})|-1 = |E|-1$, $Y(G^{c}_{-},F_{0}) = 0$,
{\emph{(ii)}}~$L(G^{x}_{-},F_{0}) = |F^{*}_{+}(G^{x}_{-},F_{0})|-1= |E|-1$, $Y(G^{x}_{-},F_{0}) = 0$.
\end{observation}
\vspace*{-0.2cm}

%% file: LinAlgebra.tex
\section{Efficient Cascading Failure Evolution Computation}\label{sec: Algebraic View}
\begin{algorithm}[t]
\caption{- Cascading Failure Evolution -- Pseudo-inverse Based (CFE-PB)}
\label{algorithm:CascadeEvol}
\vspace*{-.2cm}
\small
\begin{trivlist}
\item\textbf{Input:} A connected graph $G=(V,E)$ and an initial edge failures event $F_{0} \subseteq E$.
\end{trivlist}
\vspace*{-4mm}
\begin{algorithmic}[1]
\STATE Compute $A^+$, $F^{*}_{0} \leftarrow F_{0}$ and $i \leftarrow 0$.
\WHILE{$F_{i} \neq \emptyset$}
\FOR{each $\{r,s\}\in F_{i}$}
\IF {$\{r,s\}$ is a cut-edge (see Lemma~\ref{lem:cutedge find})}
\STATE Find the connected components after removing $\{r,s\}$. (see Lemma~\ref{lem:connected comp find})
\STATE Adjust the total demand to equal the total supply within each connected component.
\ENDIF
\STATE \textbf{else} update $A^+$ after removing $\{r,s\}$. (see Lemma~\ref{lem:update pseudo inverse})
\ENDFOR
\STATE Compute the phase angles $\hat{\Theta}=A^+P$ and compute new flows $f_{e}(F^{*}_{i})$ from the phase angles.
\STATE Find the set of new edge failures $F_{i+1}=\{e|f_e>c_e,~e\in E \setminus F^{*}_{i}\}$. $F^{*}_{i+1} \leftarrow F^{*}_{i} \cup F_{i+1}$ and $i \leftarrow i+1$.
\ENDWHILE
\RETURN $t = i-1$, $(F_{0}, \ldots, F_{t})$, and $f_{e}(F^{*}_{t})~ \forall e \in E\backslash F^{*}_{t}$.
\end{algorithmic}
\end{algorithm}
Based on the results we obtained in Section~\ref{sec:Matrixprop}, we present the Cascading Failure Evolution -- Pseudo-inverse Based (CFE-PB) Algorithm which identifies the evolution of the cascade. The CFE-PB Algorithm uses the \textit{Moore-Penrose Pseudo-inverse} of the admittance matrix for solving~(\ref{eqn:flowmatrix}). Computing the pseudo-inverse of the admittance matrix requires $O(|V|^3)$ time. However, the algorithm obtains the pseudo-inverse of the admittance matrix in round $i$ from the one obtained in round $(i-1)$, in $O(|F_i||V|^2)$ time. Moreover, in some cases, the algorithm can reuse the pseudo-inverse from the previous round.
Since once lines fail, there is a
need for low complexity algorithms to control and mitigate the cascade, the CFE-PB
Algorithm may provide insight into the design of efficient cascade control algorithms.

We now describe the CFE-PB Algorithm. It initially computes the pseudo-inverse of the admittance matrix (in $O(|V|^3)$ time) and this is the only time in which it \emph{computes $A^+$ without using a previous version of $A^+$}. Next, starting from $F_0$, at each round of the cascade, for each $e\in F_i$, it checks whether $e$ is a cut-edge (Line~4). This is done in $O(1)$ (Lemma~\ref{lem:cutedge find}). If yes, based on Lemma~\ref{lem:cutedge pseudo}, in Lines~5 and 6, the total demand is adjusted to equal the total supply within each connected component (in $O(V)$ time). Else, in Line 7, $A^+$ after the removal of $e$ is computed in $O(|V|^2)$ time (see Lemma~\ref{lem:update pseudo inverse}). After repeating this process for each $e\in F_i$, the phase angles and the flows are computed in $O(|V|^2)$ time (Line 8). The rest of the process is similar to the CFE Algorithm.

The following theorem provides the complexity of the algorithm (the proof is based on the Lemmas 1--4). We show that the algorithm runs in $O(|V|^3+|F_t^*||V|^2)$ time (compared to the CFE Algorithm which runs in $O(t|V|^3)$). Namely, if $t=|F_t^*|$ (one edge fails at each round), the CFE-PB Algorithm outperforms the CFE Algorithms by $O(\min\{|V|,t\})$.
\vspace*{-0.2cm}
\begin{theorem}\label{Th:CFEPB Complex}
CFE-PB Algorithm runs in $O(|V|^3+|F_t^*||V|^2)$ time.
\end{theorem}
\vspace*{-0.2cm}
We notice that a similar approach (the step by step rank-1 update) can also be applied to other methods for solving linear equations (e.g., LU factorization~\cite{golub2012matrix}). However, as we showed in Section~\ref{sec:Failure Analysis}, using the pseudo-inverse allows developing tools for analyzing the effect of a single edge failure. Moreover, it supports the development of an algorithm for finding the most vulnerable edges.

%% file: Hardness.tex
\section{Hardness and Heuristic}
\label{sec:hardness}
In this section, we prove that the decision problems associated with
some of the metrics are NP-complete and one of the problems is not in
APX. Using the results from Section~\ref{sec:Failure Analysis}, we introduce a heuristic algorithm for the problem of finding the set of initial failures of size $k$ that causes a cascade resulting with the minimum possible yield (\emph{minimum yield problem}). We numerically show that solutions obtained by our algorithm lead to a much lower yield than the solutions obtain by selecting the initial
edge failures randomly. Moreover, in some small graphs with a single edge failure, this algorithm obtains the optimal solution.

\subsection{Hardness}\label{subsec:hardness}
First, we show that deciding if there exists a failure
event (of size at most a given value) such that the yield after stabilization is less than a given threshold, is NP-complete.
\begin{lemma}
\label{lemma:yield-hard} Given a graph $G$, a real number $y$, $0
\leq y \leq 1$, and an integer $k \geq 1$, the problem of deciding
if $Y(G,k) \leq y$ is NP-complete.
\end{lemma}
\vspace*{-0.2cm}
We show below that deciding if there exists an initial
failure event that causes a cascade with maximum number of rounds, is
NP-complete. The proof is based on showing the relation between disconnecting a subset of supply
nodes from the graph and choosing a subset in the Partition problem. If a disconnection exists such that the total amount of
flow that reaches some central node is exactly half the total
supply (equivalent to finding
a solution for a corresponding instance of Partition problem), then
the number of rounds is strictly greater than a given threshold.
Otherwise, the number of rounds of any cascade is less than this
threshold.

\vspace*{-0.2cm}
\begin{lemma}
\label{lemma:length-hard} Given a graph $G$ and an integer $t \geq
1$, the problem of deciding if $L(G,|E|) \geq t$ is NP-complete.
\end{lemma}
\vspace*{-0.2cm}
Finally, we prove that the problem of
computing the maximum distance between consecutive edge failures is not in APX.
\vspace*{-0.2cm}
\begin{lemma}
\label{lemma:distance-hard-2}
Given a graph $G$, the problem of computing $D(G,|E|)$ is not in APX.
\end{lemma}
\vspace*{-0.2cm}
\subsection{Heuristic Algorithm for Min Yield}\label{sec:algorithm}
\begin{figure*}[t]
\vspace{-0.1cm}
\centering
\includegraphics[scale=0.28]{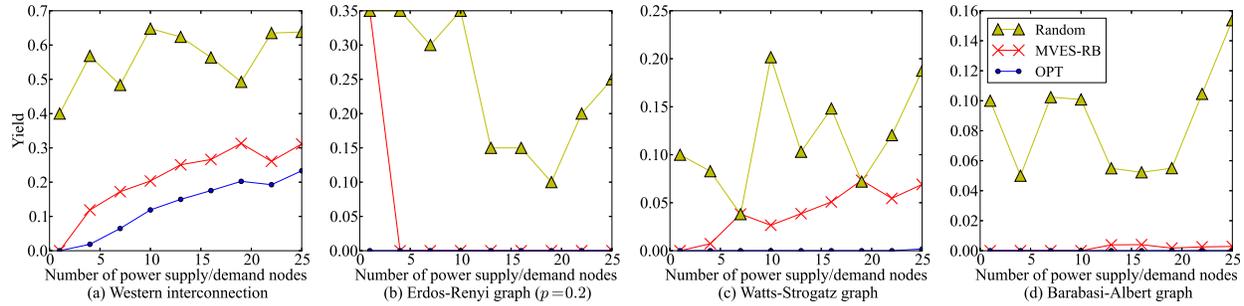}
\vspace*{-0.5cm}
\caption{The yield after stabilization when selecting a single edge failure based on the MVES-RB Algorithm, randomly, and optimally. All graphs have 136 nodes. Every data point is the average over 20 trials, each composed of a different set of supply/demand nodes.}
\vspace*{-0.4cm}
\label{figure:Alg vs opt}
\end{figure*}
\begin{algorithm}[t]
\caption{- Most Vulnerable Edges Selection -- Resistance distance Based (MVES-RB)}
\label{algorithm:kVul}
\vspace*{-.2cm}
\small
\begin{trivlist}
\item\textbf{Input:} A connected graph $G=(V,E)$ and an integer $k\geq1$.
\end{trivlist}
\vspace*{-4mm}
\begin{algorithmic}[1]
\STATE Compute $A^+$.
\STATE Compute the phase angles $\hat{\Theta}=A^+P$ and compute flows $f_{e}$ from the phase angles.
\STATE Compute the resistance distance $r(i,j)=r(e)~\forall e=\{i,j\}\in E$.
\STATE Sort edges $e_1,e_2,\dots,e_{|E|}$ such that $p\leq q$ iff $f_{e_p}r(e_p)\geq f_{e_q}r(e_q)$.
\RETURN $e_1,e_2,\dots,e_k$.
\end{algorithmic}
\end{algorithm}
\begin{figure*}[t]
\centering
\includegraphics[scale=0.28]{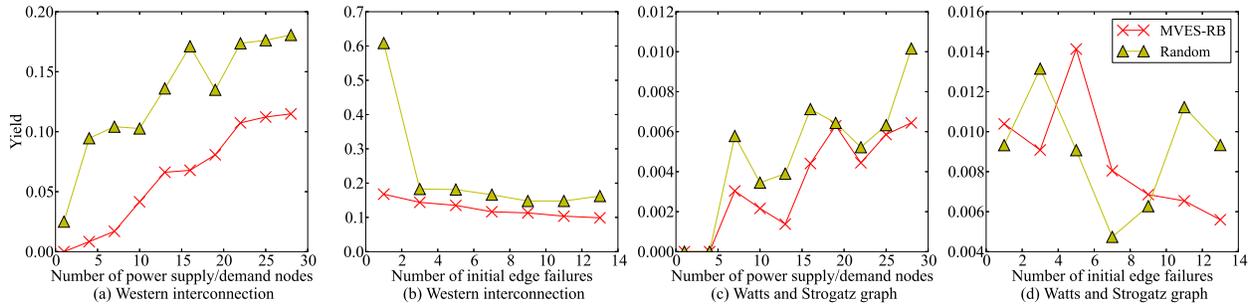}
\vspace*{-0.5cm}
\caption{The yield after stabilization when selecting an initial set of edge failures randomly and based on the MVES-RB Algorithm. In (a) and (c) number of edge failures is 10, and in (b)and (d) the number of supply/demand nodes is 20. Every data point is the average over 20 trials, each composed of a different set of supply/demand nodes.}
\label{figure:Yield Compare Num_Gen}
\vspace*{-0.4cm}
\end{figure*}
As shown in Lemma~\ref{lemma:yield-hard}, the minimum yield problem is NP-hard. We now present a heuristic algorithm for solving this problem. We refer to it as the Most Vulnerable Edge Selection -- Resistance distance Based (MVES-RB) Algorithm. From Corollary~\ref{col:upper bound1}, it seems that edges with large $r(i,j)\times |f_{ij}|$ have greater impact on the flow changes on the other edges. Based on this result, the MVES-RB Algorithm selects \emph{the $k$ edges with highest $r(i,j)\times |f_{ij}|$ values as the initial set of failures.}

The MVES-RB Algorithm is in the same category as the algorithms that identify the set of failures with the largest impact (i.e., algorithms that solve the \emph{$N-k$ problem}~\cite{Bie10,liu2013optimal,pinar_power}). However, none of the previous works focusing on the $N-k$ problem, considers cascading failures. The MVES-RB Algorithm is simpler than most of the algorithms proposed in the past. However, it is not possible to compare its performance to that of algorithms in~\cite{Bie10,liu2013optimal,pinar_power,salmeron2004analysis} since they use different formulations of the power flow problem.

We first compare via simulation the MVES-RB Algorithm to the optimal solution in small graphs and for a single initial edge failure. Fig.~\ref{figure:Alg vs opt} shows the yield after stabilization when selecting a single edge failure based on the MVES-RB Algorithm, randomly, and optimally. All the graphs have 136 nodes. For all the edges the reactance, $x_e=1$, and the capacity $c_e=1.1 f_e$,\footnote{Following~\cite{SmartGridComm11rep}, we assume that the capacities are $K$ times the initial flows on the edges. $K$ is often referred to as the \emph{Factor of Safety (FoS)} of the grid. Here, $K=1.1$ as in~\cite{SmartGridComm11rep}.} where $f_e$ is the initial flow on the edge. At each point, equal number of power supply and demand nodes are randomly selected and assigned values of 1 and -1. As can be seen, the MVES-RB Algorithm obtains the optimal solution in Erd\H{o}s-R\'{e}nyi and Bar\'{a}basi-Albert graphs. However, it does not achieve the optimal solution in the Western interconnection and Watts-Strogatz graph.

Finding the optimal solution for the minimum yield problem in general case is impossible in polynomial time. Therefore, to get better insight into the performance of the MVES-RB Algorithm, we compare it with the case that $k$ edges are selected randomly. As can be seen in Fig.~\ref{figure:Alg vs opt}, the MVES-RB Algorithm outperforms the random selection most of the time. Fig~\ref{figure:Yield Compare Num_Gen} depicts this comparison for larger initial failures in the Western interconnection and the Watts-Strogatz graph. The power supplies and demands, the reactances, and the capacities are as above. It can be seen that the MVES-RB Algorithm can perform significantly better than the random selection (Fig.~\ref{figure:Yield Compare Num_Gen}-(a) and (b)), and in some cases obtains similar performance to the random selection (Fig.~\ref{figure:Yield Compare Num_Gen}-(c) and (d)). Notice that in these cases, both methods perform relatively good (lead to yield less than 0.02).

To conclude, despite the simplicity and low complexity of the MVES-RB Algorithm, simulations indicate that it outperforms the random selection and in simple cases obtains the optimal solution.

%% file: Conclusion.tex
\section{Conclusions}\label{sec:conclusion}
We studied properties of the admittance matrix of the grid and provided analytical tools for studying the impact of a single edge failure on the flows on the other edges. Based on these tools, we derived upper bounds on the flow changes after a single edge failure and discussed the robustness of different graph classes against single edge failures. We illustrated via simulations the impact of a single edge failure. Then, we proved the unique properties of the cascading failure model and introduced a pseudo-inverse based efficient algorithm to identify the evolution of the cascade. Finally, we proved that the computational problems associated with the various metrics are hard and introduced a simple heuristic algorithm to detect the most vulnerable edges.

This is one of the first steps in using computational tools for understanding
the grid resilience to cascading failures. Hence, there are
still many open problems. In particular, we plan to study the effect of
failures on the interdependent grid and communication networks.
Moreover, while due to its relative simplicity, most
previous work in the area of grid vulnerability is based on
the DC model, this model does not capture effects such as
voltage collapse that may occur during a cascade. Hence, we
plan to develop methods to analyze the cascades
using the more realistic AC model. 

%% file: Appendix-LinearAlgebra.tex
\section{Preliminaries and Proofs of Results Used in Sections 4, 5 and 7}\label{app:prelem}

In this appendix we restate results related to the Moore-Penrose pseudo-inverse of matrix and the proofs for the results in Sections~\ref{sec:Matrixprop}, \ref{sec:Failure Analysis}, and \ref{sec: Algebraic View}.

In the following, matrices $I$ and $J$ denote the identity and the all-$1$ matrices, respectively.
\vspace*{-0.2cm}
\begin{theorem}[Moore-Penrose\cite{albert1972regression}]\label{th: pseudo inverse}
 For any $n\times m$ matrix $H$, Moore-Penrose pseudo-inverse of $H$,
\begin{eqnarray*}
H^{+}&=&\lim_{\delta \to 0} (H^t H+\delta^2 I)^{-1}H^t= \lim_{\delta \to 0} H^t (H H^t+\delta^2 I)^{-1}
\end{eqnarray*}
always exists. And for any $n$-vector $z$,
$\hat{x} = H^{+}z$
is the vector of minimum norm among those which minimize
$\|z-Hx\|$.
\end{theorem}
\vspace*{-0.2cm}

\vspace*{-0.2cm}
\begin{theorem}[Albert\cite{albert1972regression}]\label{th: pseudo invers compute}
 For any matrices $U,V$,
\begin{eqnarray*}
&&(UU^t+VV^t)^{+} = (CC^t)^+ + [I-(VC^+)^t]\\
&& \times [(UU^t)^+ - (UU^t)^+ V(I-C^+C)KV^t(UU^t)^+]\\
&& \times [1-VC^+]
\end{eqnarray*}
where $C$ and $K$ are defined as follows
\begin{equation*}
C = [I-(UU^t)(UU^t)^+]V
\end{equation*}
\begin{equation*}
K = \{I+[(I-C^+C)V^t(UU^t)^+V(I-C^+C)]\}^{-1}.
\end{equation*}
\end{theorem}
\vspace*{-0.2cm}

\vspace*{-0.2cm}
\begin{proof}[of \textbf{Lemma~\ref{lem:connected comp find}}]
Suppose that $\{i,j\}$ is a cut-edge of the connected graph $G$, and $G\backslash \{i,j\}=G_1\cup G_2$. Assume that $i\in G_1$ and $j\in G_2$. We show below that for any $\{r,s\}\in G\backslash \{i,j\}$, $a^+_{ir}-a^+_{jr}=a^+_{is}-a^+_{js}$. Moreover, for any $r\in G_1$ and $s\in G_2$, $a^+_{ir}-a^+_{jr}\neq a^+_{is}-a^+_{js}$.
Suppose that $\{r,s\}\in G\backslash \{i,j\}$ is an arbitrary edge. Then, the solution to~(\ref{eqn:flow1})-(\ref{eqn:flow2}) for the power vector $\hat{P}$ with $\hat{p}_r=-\hat{p}_s=1$ and zero elsewhere is $f_{rs}=-f_{sr}=1$ and zero elsewhere. Therefore, $f_{ij}=0$. On the other hand, from Observation~\ref{lem: pseudo inv matrix sol}, $\hat{\Theta}=A^+\hat{P}$ is a solution to the equivalent matrix equation~(\ref{eqn:flowmatrix}). Since the solution with respect to power flows is unique, $0=f_{ij}=-a_{ij}(\hat{\theta}_i-\hat{\theta}_j)= -a_{ij}(A_i^+\hat{P}-A_j^+\hat{P})\Rightarrow 0=(a_{ir}^+-a_{is}^+-a_{jr}^++a_{js}^+) \Rightarrow a_{ir}^+-a_{jr}^+=a_{is}^+-a_{js}^+$.
From this and since $a^+_{ii}-a^+_{ji}\neq a^+_{ij}-a^+_{jj}$ (Lemma~\ref{lem:cutedge find}), for any $r\in G_1$ and $s\in G_2$, $a^+_{ir}-a^+_{jr}\neq a^+_{is}-a^+_{js}$.
Thus, by using the precomputed pseudo-inverse of the admittance matrix, computing $A_i^+-A_j^+$, and dividing the entries into two groups with equal values, the connected components of $G\backslash\{i,j\}$ can be identified. This process requires $O(|V|)$ time.
\end{proof}
\vspace*{-0.2cm}
\vspace*{-0.2cm}
\begin{proof}[of \textbf{Lemma~\ref{lem:cutedge pseudo}}]
First, from Observation~\ref{lem: pseudo inv matrix sol}, $\hat{\Theta}=A^+P'$ is a solution to~(\ref{eqn:flowmatrix}) for the power vector ${P'}$ in the graph $G$. Since the solution to (\ref{eqn:flow1})-(\ref{eqn:flow2}) with respect to power flows is unique, if $f_{ij}=0$, then $\hat{\Theta}=A^+P'$ is also a solution to~(\ref{eqn:flowmatrix}) for the power vector ${P'}$ in the graph $G'$. Therefore, we only need to prove that $\hat{\theta}_{i}=\hat{\theta}_{j
}$ from $\hat{\Theta}=A^+P'$. To prove this, we prove that $\hat{\theta}_{i}-\hat{\theta}_{j}=(A^+_{i}-A^+_{j})P'=0$. However, from the proof of Lemma~\ref{lem:connected comp find}, since $\{i,j\}$ is a cut-edge, the entries of $A^+_{i}-A^+_{j}$ have equal values at the entries in the same connected component. On the other hand, since $P'$ is the power vector after load shedding, then the sum of the supplies and demands at each connected component is zero. Thus, $(A^+_{i}-A^+_{j})P'=0$.
\end{proof}
\vspace*{-0.2cm}
\vspace*{-0.2cm}
\input{lemma-pseudo-inverse}
\vspace*{-0.2cm}
\vspace*{-0.15cm}
\begin{proof}[of \textbf{Corollary~\ref{col: pseudo inv change}}]
It is easy to see from Theorem~\ref{col: pseudo inv},
\begin{equation*}\label{eq: row pseudo change}
A'^+_r=A^+_r-\frac{(a^{+}_{ri}-a^{+}_{rj})}{a_{ij}^{-1}-2(a^{+})_{ij}+(a^{+})_{ii}+(a^{+})_{jj}}(A^+_i-A^+_j).
\end{equation*}
Using this in $f_{rs}'=-a_{rs}(A'^+_r-A'^+_s)P$ completes the proof.
\end{proof}
\vspace*{-0.2cm}
\vspace*{-0.2cm}
\begin{proof}[of \textbf{Lemma~\ref{lem:update pseudo inverse}}]
Based on Corollary~\ref{col: pseudo inv change}, after the removal of a non-cut edge $\{i,j\}$, each entry of the pseudo inverse of the admittance matrix can be updated in $O(1)$ time. Thus, computing $A'^+$ from $A^+$ takes $O(|V|^2)$ time.
\end{proof}
\vspace*{-0.2cm}
\vspace*{-0.2cm}
\input{lemma-grand-flow-increase}
\vspace*{-0.2cm}
\begin{proof}[of \textbf{Corollary~\ref{col:upper bound1}}]
Using triangle inequality for resistance distance, we can write,
\begin{eqnarray*}
&&-r(i,p)+r(i,q)\leq r(p,q)\\
&&r(j,p)-r(j,q)\leq r(p,q).
\end{eqnarray*}
Apply these to Lemma~\ref{lem:flow changes} completes the proof.
\end{proof}
\vspace*{-0.2cm}
\begin{proof}[of \textbf{Corollary~\ref{col:upper bound2}}]
Notice that $r(\{i,j\},\{p,q\})=\min\{r(i,q),r(i,p),r(j,q),r(j,p)\}$. The proof is exactly the same as the proof of Corollary~\ref{col:upper bound1}.
\end{proof}
\vspace*{-0.2cm}
\vspace*{-0.2cm}
\begin{proof}[of \textbf{Observation~\ref{lem:av resisdis1}}]
From \cite[Lemma 9.9]{bapat2010graphs}, we have $\sum_{\{i,j\}\in E}r(i,j)=|V|-1$~\cite{bapat2010graphs}.
\end{proof}
\vspace*{-0.2cm}
\vspace*{-0.2cm}
\begin{proof}[of \textbf{Lemma~\ref{lem: Erdos Renyi Kf}}]
It is known that the Kirchhoff index of the graph $G$ can be written in terms of the eigenvalues of the Laplacian matrix of the graph as $Kf(G)=n\sum_{i=1}^{n-1}\frac{1}{\lambda_i}$~\cite{gutman1996quasi}. On the other hand,
\begin{equation*}
n^2\leq (\sum_{i=1}^{n-1}\frac{1}{\lambda_i})(\sum_{i=1}^{n-1}\lambda_i)=(\sum_{i=1}^{n-1}\frac{1}{\lambda_i}) \text{tr}(A).
\end{equation*}
However, when $n$ is relatively big, then each node has the degree equal to $\Theta(np)$, therefore $\text{tr}(A)=\Theta(n^2p)$. Combining this with the equations above, we can easily see that $Kf(G)=\Omega(n/p)$. Thus, the average resistance distance is of $Kf(G)/|E|=\Omega(\frac{1}{n p^2})$.

As for the upper bound, it is shown in~\cite{palacios2010bounds} that for a $d$-regular graph $H$ with $n$ nodes, $Kf(H)\leq\frac{3n^2}{d}$. Using this bound for Erd\H{o}s-R\'{e}nyi graph, we can write $Kf(G)= O(n/p)$. Thus, the average resistance distance is of $Kf(G)/|E|=O(\frac{1}{n p^2})$.
\end{proof}
\vspace*{-0.2cm}
\vspace*{-0.2cm}
\input{theorem-CFEPB-Complex}
\vspace*{-0.2cm} 

%% file: lemma-pseudo-inverse.tex
\begin{proof}[of \textbf{Theorem~\ref{col: pseudo inv}}]
First we show that if $G$ is connected, then $AA^+=I-\frac{1}{n}J$. $A$ is a real and symmetric matrix, therefore there exist an orthogonal and unitary matrix $U$ such that $A=U^tDU$, in which $D = \text{diag}(\lambda_1,\lambda_2,\dots,\lambda_n)$ is the diagonal matrix of eigenvalues of $A$ and $U_i$ is the normalized eigenvector related to eigenvalue $\lambda_i$. It is well-known that when $G$ is connected and unweighted, then the multiplicity of eigenvalue 0 of the Laplacian matrix is 1~\cite{biggs1994algebraic}. Exactly the same result with the same approach can be obtained for weighted graph, therefore we can assume that $\lambda_1=0$ and all other eigenvalues are nonzero. In this case $U_1=[\frac{1}{\sqrt{n}}, \frac{1}{\sqrt{n}}, \dots, \frac{1}{\sqrt{n}}]$. On the other hand,
 $A^+=U^tD^+U$, therefore
    \begin{eqnarray*}
    AA^+&=& U^tDUU^tD^+U=U^tDD^+U\\
    &=&U^t\text{diag}(\lambda_1\lambda_1^+, \lambda_2\lambda_2^+, \dots, \lambda_n\lambda_n^+)U\\
    &=&U^t(I-\text{diag}(1, 0, \dots, 0))U\\
    &=&I-U^t[U_1^t|0|\dots|0]^t=I-\frac{1}{n}J
    \end{eqnarray*}
    in which $[U_1^t|0|\dots|0]^t$ is an $n\times n$ matrix with $U_1$ in the first row and $0$ elsewhere.

Similarly we show that if $G$ has $k$ connected components with $m_1,m_2,\dots,m_k$ nodes, then $AA^+ = I - J_k$ in which
\begin{equation*}
J_k = \text{diag}(\frac{1}{m_1}J_{m_1\times m_1},\frac{1}{m_2}J_{m_2\times m_2},\dots,\frac{1}{m_k}J_{m_k\times m_k})
\end{equation*}
is a block matrix with matrices on the diagonal entries (with proper node indexing). Suppose $G$ has $k\leq n$ connected components. Again it is well-known that when $G$ is unweighted, multiplicity of eigenvalue 0 of the Laplacian matrix is equal to the number of connected components of graph $G$~\cite{biggs1994algebraic}. With exactly the same reasoning it can be shown that it is also the case for weighted graph. Therefore, in this case $\lambda_1=\lambda_2=\dots=\lambda_k=0$. Suppose $m_i$ is the size of the $i^{th}$ connected component. With a proper indexing of nodes, it is easy to verify that $U_i=[0,\dots,0,\frac{1}{\sqrt{m_i}},\dots,\frac{1}{\sqrt{m_i}},0,\dots,0]$, in which $u_{ij}=\frac{1}{\sqrt{m_i}}$ for $\sum_{k=1}^{i-1}m_k< j\leq\sum_{k=1}^{i}m_k$, and zero elsewhere. Now similar to previous part,
    \begin{eqnarray*}
    AA^+=I-U^t[U_1^t|U_2^t|\dots|U_k^t|0|\dots|0]^t=I-J_k.
    \end{eqnarray*}

Now we can prove the theorem. $A$ is a real and symmetric matrix, therefore there exist an $n\times n$ matrix $B$ such that $BB^t=A$. Now using Theorem~\ref{th: pseudo invers compute},
    \begin{eqnarray*}
    &&(A+a_{ij}XX^t)^{+} = (CC^t)^+ + [I-(\sqrt{a_{ij}}XC^+)^t]\\
&& \times [A^+ - a_{ij} A^+  X(I-C^+C)KX^tA^+]\\
&& \times [1-\sqrt{a_{ij}}XC^+].\footnotemark
    \end{eqnarray*}
    \footnotetext{$\sqrt{a_{ij}}$ might be an imaginary number.}
    Therefore, all we need to compute is matrices $C$ and $K$. Using previous part,
    \begin{eqnarray*}
    C = [I-AA^+]X=[I-I+J_k]X=J_k X.
    \end{eqnarray*}
    Since $\{i,j\}\in E$, nodes $i$ and $j$ should be in the same connected component of $G$. Therefore, from the structure of $J_k$, $J_k X=0$ and so $C=0$. Using this,
    \begin{eqnarray*}
    K &=& \{I+a_{ij}[(I-C^+C)X^tA^+X(I-C^+C)]\}^{-1}\\
    &=& \{I+a_{ij}[I X^tA^+X I]\}^{-1}= \{1+a_{ij}X^tA^+X\}^{-1}.
    \end{eqnarray*}
    Notice that $X$ is an $n\times 1$ vector, therefore $X^tA^+X$ is an scaler and $I$ in the second equation is $1\times1$. This is why it is written 1 instead of $I$ in the last equation. Since $\{i,j\}$ is not a cut edge, from Lemma~\ref{lem:cutedge find} we have, $1+a_{ij}X^tA^+X=a_{ij}[a_{ij}^{-1}-2(a^+)_{ij}+(a^+)_{ii}+(a^+)_{jj}]\neq0$, therefore $K$ is well-defined. Replacing $K$ and $C$,
    \begin{eqnarray*}
    &&(A+a_{ij}XX^t)^{+}  \\
    &=&A^+ - a_{ij} A^+  X\{1+a_{ij}X^tA^+X\}^{-1}X^tA^+\\
    &=& A^+ - \frac{1}{a_{ij}^{-1}+X^tA^+X} A^+  XX^tA^+
    \end{eqnarray*}
    which is what we wanted to prove.
\end{proof} 

%% file: lemma-grand-flow-increase.tex
\begin{proof}[of \textbf{Observation~\ref{lemma:grand-flow-increase}}]
We construct the graph $G=(V,E)$ as follows, $V=\{s,t\}$,
$P_s=-P_t=1$, and there are two parallel edges $e_1$ and $e_2$ between $s$ and $t$. Set the capacities $c_{e_1} =c_{e_2}=1$. Assume the reactances $x_{e_1}, x_{e_2}$ are such that $0 < x_{e_1} < x_{e_2}$.

By Eq. (\ref{eqn:flow1})-(\ref{eqn:flow2}), we get $f_{e_{1}} = \frac{x_{e_2}}{x_{e_2}+x_{e_1}}$ and $f_{e_{2}} = \frac{x_{e_1}}{x_{e_1}+x_{e_2}}$.
If $F_{0} = \{e_{1}\}$, then $f_{e_{2}}(F_{0}) = 1$ and $S_{e_2,e_1}= \frac{x_{e_2}}{x_{e_1}}$.
\end{proof} 

%% file: theorem-CFEPB-Complex.tex
\begin{proof}[of \textbf{Theorem~\ref{Th:CFEPB Complex}}]
Finding the pseudo inverse of the matrix requires $O(|V|^3)$ time. Therefore, Line~1 takes $O(|V|^3)$ time. Lines~5 and 6 in the algorithm take $O(|V|)$ time and Line~7 takes $O(|V|^2)$, therefore the whole \textbf{for} loop takes at most $O(|F_i||V|^2)$ time at each step. Using $A^+$ computed in the \textbf{for} loop, Lines~8 and 9 take $O(|V|^2)$ time. Thus, the total running time of the algorithm is at most $O(|V|^3)+O((|F_0|+|F_1|+\dots+|F_t|)|V|^2)=O(|V|^3)+O(|F_t^*||V|^2)$.
\end{proof} 

%% file: Appendix-Proofs.tex
\section{Proofs of Results from Sections 6 and 8}\label{app:proofs}

In this appendix we provide the proofs for the results in Sections 6 and 8. In our proofs, for clarity we use the notations $f_{u,v}$,  $c_{u,v}$, and $x_{u,v}$ instead of $f_{uv}$, $c_{uv}$, and $x_{uv}$ respectively. We also use the notation $\{u,v\}_1,\{u,v\}_2,\dots,etc.$ to show multiple edges between nodes $u$ and $v$.

\input{Obs-nonmonotonicity}
\input{lemma-grande-taille}
\input{lemma-grande-distance-repetition}
\input{lemma-small-change-1}

\input{lemma-yield-hard}
\input{lemma-length-hard}
\input{lemma-distance-hard-2}

%% file: Obs-nonmonotonicity.tex
\begin{proof}[\textbf{Observation~\ref{lemma:subset-1}}]
We construct the graph $G=(V,E)$ as follows,

\noindent\emph{Nodes.} $V=\{v_1,v_2,v_3,v_4\}$.

\noindent\emph{Active Powers.} $P_{v_1}=27$, $P_{v_2}=5$, $P_{v_4}=-32$, and $v_3$ is a neutral node.

\noindent\emph{Edges.} $E=\{\{v_1,v_2\},\{v_1,v_4\}_1,\{v_1,v_4\}_2,\{v_2,v_3\},\{v_3,v_4\}\}$.

\noindent\emph{Capacities.} $c_e=30$ for $e\in\{\{v_1,v_4\}_1,\{v_1,v_4\}_2\}$ and 20 \text{otherwise}.

\noindent\emph{Reactances.} All the reactances are equal to 1, except for the edge $\{v_1,v_4\}_2$, $x_{v_1,v_4}=10$.

It is easy to show that initial flows are feasible and can be computed as follows,
$f_e=2$ for $e=\{v_1,v_4\}_2$,
$f_e=5$ for $e=\{v_1,v_2\}$,
$f_e=10$ for $e\in\{\{v_2,v_3\},\{v_3,v_4\}\}$,
and $f_e=20$ for
$e=\{v_1,v_4\}_1$.

Now set $F_0=\{\{v_1,v_4\}_1\}$, the flows would change as follows,
$f_e(F_0)=7$ for
$e=\{v_1,v_4\}_2$,
$f_e=20$ for $e=\{v_1,v_2\}$,
and $f_e=25$ for $e\in\{\{v_2,v_3\},\{v_3,v_4\}\}$.
As it can be seen $f_{v_2,v_3}(F_0)$ and $f_{v_3,v_4}(F_0)$ both exceed their capacities and fail.
Therefore, $F_1=\{\{v_2,v_3\}\,\{v_3,v_4\}\}$ and flows will change as follows,
$f_e(F_1)=32$ for $e=\{v_1,v_4\}_2$, and $f_e(F_1)=5$ for $e=\{v_1,v_2\}$.
As a result, flow on the edge $\{v_1,v_4\}_2$ exceeds its capacity and fails. Following this event, the supply nodes are getting disconnected from the demand node and therefore $Y(G,F_0)=0$.

Now set $F'_0=\{\{v_1,v_4\}_1,\{v_1,v_2\}\}\subset F_0$, as the initial failure event. It is easy to show that the flows after this initial failure are feasible and have the following values, $f_e(F'_0)=27$ for $e=\{v_1,v_4\}_2$ and $f_e(F'_0)=5$ for $e\in\{\{v_2,v_3\},\{v_3,v_4\}\}$.
Thus, $Y(G,F'_0)=1$.
\end{proof} 

%% file: lemma-grande-taille.tex
\begin{proof}[\textbf{Observation~\ref{lemma:grande-taille}}]
Let $m \geq 1$ be any integer.
We construct the graph $G=(V,E)$ as follows.

\noindent
\emph{Nodes.}
Set $V=\{s,t\}$.

\noindent
\emph{Active powers.}
$P_{s}=m$~and $P_{t}=-m$.

\noindent
\emph{Edges.}
Set $E=\{\{s,t\}_{i}, 0 \leq i \leq m-1\}$.

\noindent
\emph{Capacities.}
For any $i$, $1 \leq i \leq m-1$, $c_{e_{i}}=\frac{m}{m-i}-\varepsilon$ where $\varepsilon$ is such that $0 < \varepsilon \leq \frac{1}{m-1}< \frac{m}{(m-1)(m-2)}$.
Set $c_{e_{0}} = 1$.

\noindent
\emph{Reactances.}
For any $e \in E$, $x_{e}=1$.

First note that for any $e \in E$, $f_{e} = 1$ because all the edges have equal reactance $x_{e} = 1$.
Thus the flow is feasible (for any $e \in E$, $c_{e} \geq f_e=1$) by the choice of $\varepsilon$.

Set $F_{0} = \{e_{0}\}$.
For any $e \in E \setminus F_{0}$, $f_{e}(F_{0}) = \frac{m}{m-1}$.
Thus, $f_{e_{1}}(F_{0}) > c_{e_{1}} = \frac{m}{m-1} - \varepsilon$.
Furthermore, for any $e \in E \setminus (F_{0} \cup \{e_{1}\})$,  $f_{e}(F_{0}) \leq \frac{m}{m-2} - \varepsilon \leq c_{e}$ because $0 < \varepsilon < \frac{m}{(m-1)(m-2)}$ by definition.
Thus, $F_{1} = \{e_{1}\}$.

We prove by induction that for any $i$, $1 \leq i \leq m-1$, $F_{i} = \{e_{i}\}$ .

Suppose it is true for $i$, $1 \leq i \leq m-2$, that is for any $j$, $1 \leq j \leq i$, $F_{j} = \{e_{j}\}$.
We prove that it is also true for $i+1$, that is $F_{i+1} = \{e_{i+1}\}$.
First, $f_{e_{i+1}}(F^{*}_{i})= \frac{m}{m-i-1} > c_{e_{i+1}} = \frac{m}{m-i-1} - \varepsilon$.
Furthermore, for any $e \in E \setminus (F^{*}_{i} \cup \{e_{i+1}\})$,  $f_{e}(F^{*}_{i})  \leq \frac{m}{m-i-2} - \varepsilon \leq c_{e}$ since $0 < \varepsilon < \frac{m}{(m-1)(m-2)}$ by definition.
Thus, $F_{i+1} = \{e_{i+1}\}$. As a result, $L(G,1) = |E| - 1$, $|F^{*}_{+}(G,1)|=|E|$, and $Y(G,1)=0$.
\end{proof} 

%% file: lemma-grande-distance-repetition.tex
\begin{proof}[\textbf{Observation~\ref{lemma:grande-distance-repetition}}]
Without loss of generality we can assume $d$ is odd, otherwise we can proof the Observation for $d+1$.
Choose $q = \frac{d-1}{2}$.
We construct $G=(V,E)$ as follows.

\noindent
\emph{Nodes.}
Set $V = S \cup U \cup T$ where $S = \{s\}$, $U = \{u^{i}_{j}, 0 \leq i \leq l-1, 1 \leq j \leq d-2\}$,
$T = \{t\}$.

\noindent
\emph{Active powers.}
$P_{s} = l$, $P_{t} = -l$, and $P_{u} = 0~\forall u \in U$.

\noindent
\emph{Edges.}
Set $E = SU \cup UU \cup UT \cup ST$ where
$SU = \{\{s,u^{i}_{1}\}, 0 \leq i \leq l-1\}$,
$UU = \{\{u^{i}_{j},u^{i}_{j+1}\}, 0 \leq i \leq l-1, 1 \leq j \leq d-3\}$,
$UT = \{\{u^{i}_{d-2},t\}, 0 \leq i \leq l-1\}$,
and $ST = \{\{s,t\}\}$.

\noindent
\emph{Capacities.} $c_e=\frac{l}{l-i} - \varepsilon$ for $e=\{u^{i}_{q},u^{i}_{q+1}\}~ \forall 0\leq i\leq l-1$ and
$c_e=l$  \text{otherwise}, where $\varepsilon$ is such that $0 < \varepsilon < \frac{1}{l-1}$.

\noindent
\emph{Reactances.} $x_{e} =\mu$ for $e \in ST$ and 1 for $e \in E \setminus ST$, where $\mu$ is such that $0 < \mu < \frac{d - \varepsilon d}{\varepsilon l}$.

First, it is easy to see that initial flows are feasible and can be computed as follows, $f_e=\frac{ld}{\mu l + d}$ for $e=\{s,t\}$, and $f_e=\frac{\mu l}{\mu l + d}$ for $e \in E \setminus ST$.

\vspace*{-0.2cm}
\begin{claim}
\label{claim:distance-1}
If $F_{0} = ST$, then $F_{i} = \{\{u^{i-1}_{q},u^{i-1}_{q+1}\}\},~\forall i:~1 \leq i \leq l$.
\end{claim}
\vspace*{-0.2cm}
\vspace*{-0.2cm}
\begin{proof}
By strong induction on $i$. By construction of $G$, for any $e \in E \setminus F_{0}$, $f_{e}(F_{0}) = 1$.
Thus, $f_{u^{0}_{q},u^{0}_{q+1}} > c_{u^{0}_{q},u^{0}_{q+1}} = 1 - \varepsilon$.
But for any $e \in E \setminus (F_{0} \cup \{u^{0}_{q},u^{0}_{q+1}\})$, $f_{e} \leq c_{e}$ because $c_{e} \geq \frac{l}{l-1}-\varepsilon \geq 1$.
Therefore, $F_{1} = \{u^{0}_{q},u^{0}_{q+1}\}$.

Now suppose the claim is true for $1 \leq i \leq l-1$, that is $F_{i} = \{\{u^{i-1}_{q},u^{i-1}_{q+1}\}\}$.
By induction hypothesis, for any $e \in E \setminus F^{*}_{i}$, $f_{e}(F^{*}_{i}) \in \{0, \frac{l}{l-i}\}$, in particular $f_{u^{i}_{q},u^{i}_{q+1}}= \frac{l}{l-i}$.
Thus, $f_{u^{i}_{q},u^{i}_{q+1}} > c_{u^{i}_{q},u^{i}_{q+1}} = \frac{l}{l-i} - \varepsilon$.
Moreover, for any $e \in E \setminus (F^{*}_{i}\cup \{\{u^{i}_{q},u^{i}_{q+1}\}\})$, $f_{e} \leq c_{e}$ because $c_{e} \geq \frac{l}{l-(i+1)}-\varepsilon \geq \frac{l}{l-i}$.
Therefore, $F_{i+1} = \{\{u^{i}_{q},u^{i}_{q+1}\}\}$.
\end{proof}
\vspace*{-0.2cm}
\vspace*{-0.3cm}
\begin{claim}
\label{claim:la-longueur}
$L(G,1) = l$.
\end{claim}
\vspace*{-0.3cm}
\vspace*{-0.2cm}
\begin{proof}
By definition, $L(G,0)=0$.
Recall that for any $F \subseteq E$ and $e \in E$, $f_{e}(F) \leq l$.
Thus, $L(G) \leq l$ because $|\{e: c_{e} < l, e \in E\}| = l$.
In another hand by Claim~\ref{claim:distance-1}, $L(G,1) \geq l$.
\end{proof}
\vspace*{-0.2cm}
\vspace*{-0.3cm}
\begin{claim}
\label{claim:distance-grande}
If $F_{0} = ST$, then for any $i$, $1 \leq i \leq l$, $d_{i} \geq d$.
\end{claim}
\vspace*{-0.3cm}
\vspace*{-0.2cm}
\begin{proof}
By Claim~\ref{claim:distance-1}, $F_{i} = \{\{u^{i-1}_{q},u^{i-1}_{q+1}\}\}$ $\forall i: 1 \leq i \leq l$.
By construction of $G$, for any $i,j$, $1 \leq i,j \leq l$, $i \neq j$,
$d(F_{i},F_{j}) = d$.
As $d \geq d$, for any $i$, $1 \leq i \leq L(G,1)=t$, $d_{i} \geq d$.
\end{proof}
\vspace*{-0.2cm}
Claims \ref{claim:la-longueur} and \ref{claim:distance-grande} complete the proof.
\end{proof} 

%% file: lemma-small-change-1.tex
\begin{proof}[\textbf{Observation~\ref{lemma:small-change-1}}]
Let $m \geq 1$ be any integer.
We construct the graph $G=(V,E)$ as follows.
(Graphs $G^{c}_{-}$ and $G^{x}_{-}$ are identical to $G$ except in the capacity and the reactance of an edge repectively.
These slight modifications are detailed in the description of $G$.)

\noindent
\emph{Nodes.}
Set $V=\{s,t\}$.

\noindent
\emph{Active powers.} $P_s=-P_t=m$.

\noindent
\emph{Edges.}
Let $E=\{\{s,t\}_{i}, 0 \leq i \leq m-1\}$. For convenience, we show the edge $\{s,t\}_{i}$ by $e_i$.

\noindent
\emph{Capacities.} In $G$, $c_{e_i}=1$ for $i=0$, $c_e=\frac{m}{m-1}$ for $i=1$, and $c_e=\frac{m}{m-i}-\varepsilon$ for $2 \leq i \leq m-1$, where $\varepsilon$ is such that $0 < \varepsilon \leq \frac{1}{m-1} < \frac{m}{(m-1)(m-2)}$.

In $G^{c}_{-}$, $c^{-}_{e_i}=1$ for $i=0$, and $c^{-}_{e_i}=\frac{m}{m-i}-\varepsilon$ for $1 \leq i \leq m-1$.

\noindent
\emph{Reactances.}
In $G$, for any $e \in E$, $x_{e}=1$.
In $G^{x}_{-}$, for any $e \in E \setminus \{e_{1}\}$, $x^{-}_{e} = 1$, and $x^{-}_{e_{1}} = 1-\mu$, $0 < \mu \leq \frac{1}{m-1} $.

Now set $F_{0} = \{\{s,t\}_{0}\}$. In $G$, for any $e \in E \setminus F_{0}$, $f_{e}(F_{0}) = \frac{m}{m-1} \leq c_{e}$.
Thus, $L(G,F_{0}) = 0$, $|F^{*}_{+}(G,F_{0})| = |F_{0}| = 1$, and $Y(G,F_{0}) = 1$.

\emph{a)}
Consider the graph $G^{c}_{-}$.
Note that the graph used in the proof of Observation~\ref{lemma:grande-taille} is exactly $G^{c}_{-}$.
We deduce that for any $e \in E$, $f_{e} \leq c_{e}$, and so the flow is feasible.
Furthermore, $L(G^{c}_{-},F_{0}) = |F^{*}_{+}(G^{c}_{-},F_{0})|-1 = |E|-1$ and $Y(G^{c}_{-},F_{0}) = 0$.

\emph{b)}
Consider the graph $G^{x}_{-}$.
$\mu \leq \frac{1}{m-1}$, so $f_{e_{1}} \leq \frac{m}{m-1} = c_{e_{1}}$, in the other hand for any $e \in E \setminus \{e_{1}\}$, $f_{e} = (1 - \mu) f_{e_{1}}$, and so $f_{e} \leq c_{e}$.
Thus the flow is feasible in $G^{x}_{-}$.
Now, since $\mu > 0$, $f_{e_{1}}(F_{0}) > \frac{m}{m-1} = c_{e_{1}}$.
Furthermore, for any $e \in E \setminus (F_{0} \cup \{e_{1}\})$, $f_{e}(F_{0}) \leq \frac{m}{m-2}-\varepsilon = c_{e}$ by the choice of $\varepsilon$ and $\mu$.
Thus $F^{*}_{1} = \{e_{0}, e_{1}\}$.
Observe that the graph $(V,E \setminus F^{*}_{1})$ is exactly the graph used in the proof of Observation~\ref{lemma:grande-taille} when removing edges $e_{0}$ and $e_{1}$.
Recall that the difference between $G$ and $G^{x}_{-}$ is only the reactance of edge $e_{1}$, that has been removed.
Thus, we get $L(G^{x}_{-},F_{0}) = |F^{*}_{+}(G^{x}_{-},F_{0})|-1= |E|-1$, and $Y(G^{x}_{-},F_{0}) = 0$.
\end{proof}

%% file: lemma-yield-hard.tex
\begin{proof}[\textbf{Lemma~\ref{lemma:yield-hard}}]
Consider following problem:
\vspace*{-0.2cm}
\begin{problem}\label{prob: multicut NP}
Suppose $G=(V,E)$ is an instance of the classical flow problem, with a single source node $\{s\}$ and set of sink nodes $T$. Assume demands are equal to 1 and lines have unbounded capacity ($O(|V|)$). Does a subset of edges $\mathcal{A}\subseteq E$ with $|\mathcal{A}|\leq k$ exist such that $|T_{fail}|\geq m$? ($T_{fail}$ is set of sink nodes which get disconnected from the source node $s$ after removing set of edges $\mathcal{A}$.)
\end{problem}
\vspace*{-0.2cm}
It is proved in \cite[Theorem 7]{aura2000analyzing}, that problem \ref{prob: multicut NP} is NP-complete. We want to use this result to proof Lemma \ref{lemma:yield-hard}. For this reason we provide a polynomial time reduction from problem above to minimum yield problem.
\vspace*{-0.2cm}
\begin{problem}\label{prob: yield NP}
Suppose $G=(V,E)$ is an instance of the power flow problem, with set of supply node $S=\{s\}$ and set of demand nodes $T$. Assume $P_t=-1$ for all $t\in T$, and $P_s=|T|$. Assume all the lines have  capacities equal to $|T|$ and reactances equal to 1. Is $Y(G,k)\leq 1-\frac{m}{|T|}$?
\end{problem}
\vspace*{-0.2cm}
\vspace*{-0.2cm}
\begin{claim}\label{claim: NP-red}
Suppose the graphs in problems \ref{prob: multicut NP} and \ref{prob: yield NP} are the same, then the answer to problem \ref{prob: multicut NP} is \emph{yes} if, and only if, the answer to problem \ref{prob: yield NP} is \emph{yes}.
\end{claim}
\vspace*{-0.2cm}
\vspace*{-0.2cm}
\begin{proof}
($\Rightarrow$) Assume the answer to problem \ref{prob: multicut NP} is \emph{yes}. It means that there exists a set of edges $\mathcal{A}\subseteq E$ with $|\mathcal{A}|\leq k$ such that their removal disconnects at least $m$ of the sink nodes from the source node. Now in problem \ref{prob: yield NP}, choose $F_0=\mathcal{A}$. Since two graphs are the same, at least $m$ of the demand nodes are disconnected from the supply node $s$. As a result, final yield is at most $|T|-m$. Since initial yield was $|T|$, $Y(G,F_0)\leq 1-\frac{m}{|T|}$. Hence, $Y(G,k)\leq 1-\frac{m}{|T|}$.

\noindent($\Leftarrow$) Now the other way, assume the answer to problem \ref{prob: yield NP} is \emph{yes}. It means that there is an initial set of edge failures $F_0\subseteq E$ with $|F_0|\leq k$ such that $Y(G,F_0)\leq 1-\frac{m}{|T|}$. First, since all the edges have capacity equal to $|T|$ which is an upper bound for a flow in an edge, after initial set of failures, there is no cascade. Therefore, there is no further edge failures. Second, with the same reason, as long as a demand node is connected to the supply node, its demand can be satisfied. Now since $Y(G,F_0)\leq 1-\frac{m}{|T|}$,  with initial set of failure $F_0$, at least $m$ of the demand nodes are disconnected from supply node $s$. In problem \ref{prob: multicut NP} choose $\mathcal{A}=F_0$, since the graphs in two problems are the same, by removing set of edges $\mathcal{A}$ from $G$, at least $m$ of the sink nodes are disconnected from source node $s$. Since $|\mathcal{A}|=|F_0|\leq k$, the answer to problem \ref{prob: multicut NP} is also \emph{yes}.
\end{proof}
\vspace*{-0.2cm}
It can be concluded from this claim that problems \ref{prob: multicut NP} and \ref{prob: yield NP} are equivalent. Therefore, problem \ref{prob: yield NP} is also NP-complete. Now since \ref{prob: yield NP} is an special case of the minimum yield problem, the minimum yield problem is NP-hard, and hence its decision version is NP-complete.
\end{proof} 

%% file: lemma-length-hard.tex
\begin{proof}[\textbf{Lemma~\ref{lemma:length-hard}}]
Let $A = \{a_{i}, 1 \leq i \leq n\}$ be an instance of the Partition problem such that $\sum_{i=1}^{n} a_{i} = 2m$.
Let $L \leq \min\{m,n\}$ be any integer.
We construct the graph $G=(V,E)$ as follows.

\noindent
\emph{Nodes.}
Set $V = S \cup S' \cup U \cup T$ where
$S = \{s_{i}, 1 \leq i \leq n\}$,
$S' = \{s',s''\}$,
$U = \{u^{i}_{j}, 0 \leq i \leq L-1, 1 \leq j \leq L-1\}$,
and $T = \{t\}$.

\noindent
\emph{Active powers.}
$P_{s_{i}} = a_{i}$ for $1 \leq i \leq n$, and $P_{t} = -2m$.
The other nodes are neutral.

\noindent
\emph{Edges.}
Set $E = SS' \cup S'S'' \cup S''U \cup UU \cup UT \cup S'T$ where
{$SS' = \{\{s_{i},s'\}, 1 \leq i \leq n\}$,}
{$S'S'' = \{s',s''\}$,}
{$S''U = \{\{s'',u^{i}_{1}\}, 0 \leq i \leq L-1\}$,}
{$UU = \{\{u^{i}_{j},u^{i}_{j+1}\}, 0 \leq i \leq L-1, 1 \leq j \leq L-3\}$,}
{$UT = \{\{u^{i}_{L-1},t\}, 0 \leq i \leq L-1\}$,}
and {$S'T = \{s',t\}$.}

\noindent
\emph{Capacities.}
\begin{equation*}
c_e=
\begin{cases}
 m & e \in S'S'', \\
 2m & e \in SS' \cup S'T\cup S''U\cup UT, \\
\frac{m}{L-i} - \varepsilon & e \in \{\{u^{i}_{j},u^{i}_{j+1}\}, 1 \leq j \leq L-2\}
\end{cases}
\end{equation*}
where $\varepsilon$ is such that $0 < \varepsilon < \frac{1}{m-1}$.

\noindent
\emph{Reactances.} $x_e=1~\forall~ e\in E$.

\begin{figure}[t]
\begin{center}
\includegraphics[scale=0.8]{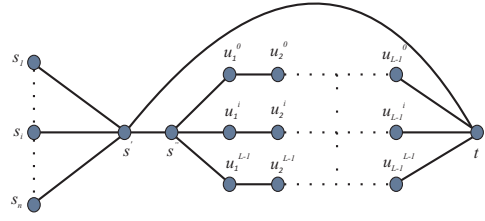}
\end{center}
\vspace{-0.5cm}
\caption{Graph $G$ described in the proof of Lemma~\ref{lemma:length-hard}.}
\label{figure:length-hard}
\end{figure}
The graph $G$ is depicted in Fig.~\ref{figure:length-hard}.

Using $G$, the proof would be as follows.
First, we show in Claim~\ref{claim:petit-a-1} that if $\{s',s''\} \in F_{0}$, then there is no cascading failures.
In other words, $F^{*} = F_{0}$.
Claim~\ref{claim:petit-a-2} proves that if at least one edge $e \in S''U \cup UU \cup UT$ belongs to $F_{0}$, then the number of rounds is at most $L-1$.
Claim~\ref{claim:petit-a-3} shows that is $\{s',t\} \in F_{0}$, then the number of rounds of the cascade is at most $L-1$.
As a corollary of Claims~\ref{claim:petit-a-1},~\ref{claim:petit-a-2}, and~\ref{claim:petit-a-3}, Claim~\ref{claim:corollary-a} shows three necessary conditions to have a number of rounds at least $L$.
Finally Claims~\ref{claim:length-hard-last-1} and~\ref{claim:length-hard-last-2} prove that the number of rounds is at least $L$ if, and only if, there exists a solution for the instance $A = \{a_{i}, 1 \leq i \leq n\}$ of the Partition problem.
In other words, the number of rounds is at least $L$ if, and only if, there exists $F_{0}$ such that the flow that go through node $s'$ is exactly $m$.

First it is easy to see that initial flows are feasible and have the following values,
\begin{equation*}
f_e=
\begin{cases}
P_{s_{i}}  & e=\{s_{i},s'\}, 1\leq i\leq n,\\
\frac{4m}{3}  & e=\{s',t\},\\
\frac{2m}{3}  & e=\{s',s''\}, \\
\frac{2m}{3L} & e \in S''U \cup UU \cup UT.
\end{cases}
\end{equation*}

\vspace*{-0.23cm}
\begin{claim}
\label{claim:petit-a-1}
If $\{s',s''\} \in F_{0}$, then $F^{*} = F_{0}$.
\end{claim}
\vspace*{-0.23cm}
\vspace*{-0.25
cm}
\begin{proof}
Suppose $\{s',s''\} \in F_{0}$.
For any $e \in S''U \cup UU \cup UT$, if $e \notin F$, then $e \notin F^{*}$.
Indeed, by construction of $G$, for any $e \in S''U \cup UU \cup UT$, $f_{e}(F_{0}) = 0$.
Furthermore, for any $e \in SS' \cup S'T$, if $e \notin F_{0}$, then $e \notin F^{*}$ because $c_{e} = 2m$ for any $e \in SS' \cup S'T$.
\end{proof}
\vspace*{-0.25
cm}
\vspace*{-0.23cm}
\begin{claim}
\label{claim:petit-a-2}
Let $F_{0} \subseteq E$ be a set of initial edge failures.
If $F_{0} \cap (S''U \cup UU \cup UT) \neq \emptyset$, then $L(G,F_{0}) \leq L-1$.
\end{claim}
\vspace*{-0.23cm}
\vspace*{-0.23cm}
\begin{proof}
For any $i$, $0 \leq i \leq L-1$, let $E_{i} = \{s'',u^{i}_{1}\} \cup \{\{u^{i}_{j},u^{i}_{j+1}\}, 1 \leq j \leq L-2\} \cup \{u^{i}_{L-1},t\}$.
Let $e \in F_{0} \cap E_{i}$ for some $i$, $1 \leq i \leq L-1$.
By construction, for any $e' \in E_{i}$, $f_{e'}(F_{0}) = 0$.
Thus, for any $j$, $1 \leq j \leq t$, $E_{i} \cap F_{j} = \emptyset$, where  $t$ is the number of rounds.

Suppose now that $t \geq L$.
It means that there exists $i$, $1 \leq i \leq L-1$, such that $e, e' \in E_{i}$ and $e \in F_{j_{1}}$ and $e' \in F_{j_{2}}$ with $0 \leq j_{1} < j_{2} \leq L-1$.
A contradiction because by construction $f_{e'}(F^{*}_{j_{1}}) = 0$ and $ j_{1} < j_{2}$.
Thus, $e' \notin F_{j_{2}}$ and $t \leq L-1$.
\end{proof}
\vspace*{-0.23cm}
\vspace*{-0.23cm}
\begin{claim}
\label{claim:petit-a-3}
If $\{s',t\} \notin F_{0}$, then $L(G,F_{0}) \leq L-1$.
\end{claim}
\vspace*{-0.23cm}
\vspace*{-0.23cm}
\begin{proof}
By contradiction. Assume $\{s',t\} \notin F_{0}$, and $L(G,F_{0}) \geq L-1$.
By Claim~\ref{claim:petit-a-1}, $\{s',s''\} \notin F_{0}$. Otherwise we would get $L(G,F_{0}) \leq L-1$.
By Claim~\ref{claim:petit-a-2} , $F_{0} \cap (S''U \cup UU \cup UT) = \emptyset$.
Again, otherwise we would get $L(G,F_{0}) \leq L-1$. Thus $F_0\subseteq SS'$ and
by initial flow values, we get:
\begin{itemize}
\item $f_{s_{i},s'}(F_0)  = P_{s_{i}} \leq c_{s_{i},s'} \forall~\{s_{i},s'\}\in SS'\backslash F_0$ ;
\item $f_{s',t}(F_0) \leq \frac{4m}{3} \leq c_{s',t}$;
\item $f_{s',s''}(F_0) \leq \frac{2m}{3} \leq c_{s',s''}$;
\item For any $e \in S''U(F_0) \cup UU \cup UT$, $f_{e} \leq \frac{2m}{3L} \leq c_{e}$.
\end{itemize}
Therefore, no link exceeds its capacity and $L(G,F_{0})=0$ which is a contradiction with our initial assumption.
Thus, if $\{s',t\} \notin F_{0}$, then $L(G,F_{0}) \leq L-1$.
\end{proof}
\vspace*{-0.23cm}
\vspace*{-0.23cm}
\begin{claim}
\label{claim:corollary-a}
Let $F_{0} \subseteq E$ be a set of initial edge failures.
If $L(G,F_{0}) \geq L$, then $F_{0} \cap (S''U \cup UU \cup UT) = \emptyset$, $\{s',t\} \in F_{0}$, and $\{s',s''\} \notin F_{0}$.
\end{claim}
\vspace*{-0.23cm}
\vspace*{-0.23cm}
\begin{proof}
Claim~\ref{claim:petit-a-1}, Claim~\ref{claim:petit-a-2}, and Claim~\ref{claim:petit-a-3} prove the result.
\end{proof}
\vspace*{-0.23cm}
\vspace*{-0.23cm}
\begin{claim}
\label{claim:length-hard-last-1}
If there exists a solution for the instance $A = \{a_{i}, 1 \leq i \leq n\}$ of the Partition problem, then there exists a set of initial edge failures $F_{0} \subseteq E$ such that $L(G,F_{0}) = L$.
\end{claim}
\vspace*{-0.23cm}
\vspace*{-0.23cm}
\begin{proof}
Suppose there exists a solution for the instance $A = \{a_{i}, 1 \leq i \leq n\}$ of the Partition problem, that is there exists a subset $H \subset A$ such that $\sum_{i=1,a_{i} \in H}^{n} a_{i} = m$.
We set $F(SS') = \{\{s_{i},s'\}, a_{i} \in H\}$.
By Claim~\ref{claim:corollary-a}, we set $F_{0} = F(SS') \cup S'T$.
Indeed, otherwise we would get a cascade with a number of rounds $t \leq L-1$.
Clearly, we have $f_{s',s''}(F_{0}) = m = c_{s',s''}$.
For any $i$, $0 \leq i \leq L-1$, let $E_{i} = \{s'',u^{i}_{1}\} \cup \{\{u^{i}_{j},u^{i}_{j+1}\}, 1 \leq j \leq L-2\} \cup \{u^{i}_{L-1},t\}$.
We prove that for any $i$, $1 \leq i \leq L$, $F_{i} = E_{i-1}$.
By induction on $i$

\begin{itemize}
\item $i=1$: \emph{Computation of $f_{e}(F_{0})$ and $F_{1}$ based on $F_{0}$.}

Using Eq. (\ref{eqn:flow1}) and (\ref{eqn:flow2}), for any $e \in S''U \cup UU \cup UT$, $f_{e}(F_{0}) = \frac{m}{L}$ because $\{s',t\} \in F_{0}$ and $e \notin F_{0}$.

For any $e \in E_{0}$, $f_{e}(F_{0}) = \frac{m}{L} > c_{e} = \frac{m}{L} - \varepsilon$.
Furthermore, for any $e \in (S''U \cup UU \cup UT) \setminus E_{0}$, we get $f_{e}(F_{0}) = \frac{m}{L} \leq c_{e} \leq \frac{m}{L-1} - \varepsilon$ because $0 < \varepsilon < \frac{1}{m-1}$.
Thus $F_{1} = E_{0}$.

\item Suppose it is true for $i$, $1 \leq i \leq L-1$, then we prove it is also true for $i+1$. \\
\emph{Computation of $f_{e}(F^{*}_{i})$ and $F_{i+1}$ based on $F^{*}_{i}$.}

For any $j$, $1 \leq j \leq i$, suppose $F_{j} = E_{j-1}$ .
We prove that $F_{i} = E_{i-1}$.

By Eq. (\ref{eqn:flow1}) and (\ref{eqn:flow2}), for any $e \in (S''U \cup UU \cup UT) \setminus F^{*}_{i}$, $f_{e}(F^{*}_{i}) = \frac{m}{m-i}$ because $\{s',t\} \in F^{*}_{i}$.

For any $e \in E_{i}$, we get $f_{e}(F^{*}_{i}) = \frac{m}{L-i} > c_{e} = \frac{m}{L-i} - \varepsilon$.
Furthermore, for any $e \in (S''U \cup UU \cup UT) \setminus (F^{*}_{i} \cup E_{i})$, we get $f_{e}(F^{*}_{i}) = \frac{m}{L-i} \leq c_{e} \leq \frac{m}{L-i-1} - \varepsilon$ because $0 < \varepsilon < \frac{1}{m-1}$.
Thus $F_{i+1} = E_{i}$.
\end{itemize}

Finally, $L(G,F_{0}) = L$.
\end{proof}
\vspace*{-0.23cm}
\vspace*{-0.23cm}
\begin{claim}
\label{claim:length-hard-last-2}
If there does not exist a solution for the instance $A = \{a_{i}, 1 \leq i \leq n\}$ of the Partition problem, then any set of initial edge failures $F_{0} \subseteq E$ is such that $L(G,F_{0}) \leq L-1$.
\end{claim}
\vspace*{-0.23cm}
\vspace*{-0.23cm}
\begin{proof}
Suppose there does not exist a solution for the instance $A = \{a_{i}, 1 \leq i \leq n\}$ of the Partition problem, that is for any subset $H \subset S$, $\sum_{i=1,a_{i} \in H}^{n} a_{i} \neq m$.
For such a subset, we set $F(SS') = \{\{s_{i},s'\}, a_{i} \in H\}$.

By contradiction.
Suppose there exists a set of initial edge failures $F_{0} \subseteq E$ such that $L(G,F_{0}) \geq L$.

By Claim~\ref{claim:corollary-a}, we set $F_{0} = F(SS') \cup S'T$.
Indeed, otherwise we would get a cascade length of size at most $t \leq L-1$.

Let $f' = f_{s',s''}(F_{0}) \neq m$.
There are two cases: (i) if $f' \geq m+1$, then $f_{s',s''}(F_{0}) \geq m+1 > c_{s',s''} = m$.
Thus, $\{s',s''\} \in F_{1}$ and the length of the cascade is $t = 1$ because the supply nodes are not connected anymore to the demand node, (ii) If $f' \leq m-1$, then for any $e \in S''U \cup UU \cup UT$, $f_{e}(F_{0}) = \frac{f'}{L} \leq \frac{m-1}{L} \leq \frac{m}{L} - \varepsilon \leq c_{e}$. Thus, $F^{*} = F_{0}$ and the length of the cascade is $t = 0$.
\end{proof}
\vspace*{-0.23cm}
Finally, there exists a set of initial edge failures $F_{0} \subseteq E$ such that $L(G,F_{0}) = L$ if and only there exists a solution for the instance $A$ of the Partition problem.
Thus, $L(G,|E|) = L$ if and only there exists a solution for the instance $A$ of Partition problem.
Furthermore, our reduction is polynomial.
As the Partition problem is NP-complete~\cite{Kar72}, then the decision problem associated with $L(G,|E|)$ is NP-complete.
\end{proof}

%% file: lemma-distance-hard-2.tex

\begin{proof}[\textbf{Lemma~\ref{lemma:distance-hard-2}}]

Let $d$ be an even integer.
Let $A = \{a_{i}, 1 \leq i \leq n\}$ be an instance of the Partition problem such that $\sum_{i=1}^{n} a_{i} = 2m$.
We construct the graph $G=(V,E)$ as follows.

\noindent
\emph{Nodes.}
Set $V = S \cup U \cup T$ where $S = \{s_{i}, 1 \leq i \leq n\}$,
$U = \{u_{j}, 0 \leq j \leq d\}$,
$T = \{u_{d+1}\}$.

\noindent
\emph{Active powers.}
$P_{s_{i}} = a_{i}$ for $1 \leq i \leq n$, and $P_{u_{d+1}} = -2m$.
The other nodes are neutral.

\noindent
\emph{Edges.}
Set $E = SU \cup UU$ where $SU = \{\{s_{i},u_{0}\}, 1 \leq i \leq n\}$ and
$UU = \{\{u_{j},u_{j+1}\}, 0 \leq j \leq d\} \cup \{u_{0},u_{d+1}\}$.

\noindent
\emph{Capacities.} $c_e=m$ for $e=\{u_{0},u_{1}\}$,
$c_e=m - \varepsilon$ for $e=\{u_{\frac{d}{2}},u_{\frac{d}{2}+1}\}$ (where $\varepsilon$ is such that $0 < \varepsilon < 1$.
), and
$c_e=2m$ otherwise.

\noindent
\emph{Reactances.} $x_e=\mu$ for $e=\{u_{0},u_{d+1}\}$,
where $\mu$ is such that $0 < \mu < \frac{(d+1)(m-\varepsilon)}{m+\varepsilon}$, and $x_e=1$ otherwise.

\begin{figure}[t]
\begin{center}
\includegraphics[scale=0.7]{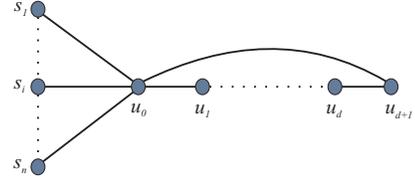}
\end{center}
\vspace{-0.5cm}
\caption{Graph $G$ described in the proof of Lemma~\ref{lemma:distance-hard-2}.}
\label{figure:distance-hard}
\end{figure}

The graph $G$ is depicted in Figure~\ref{figure:distance-hard}.

Using $G$, the proof would be as follows.
 First, we prove useful Claims~\ref{claim:label-D1},~\ref{claim:label-D2},~\ref{claim:label-D3} in order to show in Claim~\ref{claim:label-D4} that $D(G,|E|) \in \{0,d\}$.
Finally Claims~\ref{claim:label-D5} and~\ref{claim:label-D6} prove that the distance between edge failures is $d$ if, and only if, there exists a solution for the instance $A = \{a_{i}, 1 \leq i \leq n\}$ of the Partition problem.
We conclude that the problem of computing the maximum possible distance between consecutive failures cannot be approximated within any constant in polynomial time, unless P=NP.
In other words, the problem is not APX.

First, it is easy to see that initial flows are feasible and can be computed as follows,
\begin{equation*}
f_e=
\begin{cases}
P_{s_{i}} & e=\{s_{i},u_{0}\}~\text{for}~1\leq i\leq n,\\
\frac{2m(d+1)}{\mu + d + 1} & e=\{u_{0},u_{d+1}\},\\
\frac{2 \mu m}{\mu + d + 1} & \text{otherwise}.
\end{cases}
\end{equation*}
\vspace*{-0.2cm}
\begin{claim}
\label{claim:label-D1}
For any set $F_{0} \subseteq E$ and for any $e \in E \setminus \{\{u_{0},u_{1}\},\{u_{\frac{d}{2}},u_{\frac{d}{2}+1}\}\}$, if $e \notin F_{0}$, then $e \notin F^{*}$.
\end{claim}
\vspace*{-0.2cm}
\vspace*{-0.2cm}
\begin{proof}
For any $e \in E$, $f_{e}(F) \leq 2m$.
Thus, for any $e \in E \setminus \{\{u_{0},u_{1}\},\{u_{\frac{d}{2}},u_{\frac{d}{2}+1}\}\}$, if $e \notin F_{0}$, then $e \notin F^{*}$ because $c_{e} = 2m$.
\end{proof}
\vspace*{-0.2cm}
\vspace*{-0.2cm}
\begin{claim}
\label{claim:label-D2}
For any set $F_{0} \subseteq E$, $F^{*} \setminus F_{0}$ is either equal to $\emptyset$ or $\{u_{0},u_{1}\}$, or $\{u_{\frac{d}{2}},u_{\frac{d}{2}+1}\}$.
\end{claim}
\vspace*{-0.2cm}
\vspace*{-0.2cm}
\begin{proof}
By Claim~\ref{claim:label-D1}, $F^{*} \setminus F_{0} \subseteq \{\{u_{0},u_{1}\},\{u_{\frac{d}{2}},u_{\frac{d}{2}+1}\}\}$.
Now if $\{u_{0},u_{1}\}$ fails at some point before $\{u_{\frac{d}{2}},u_{\frac{d}{2}+1}\}$, then it prevents $\{u_{\frac{d}{2}},u_{\frac{d}{2}+1}\}$ to fail due to failure model. If $\{u_{\frac{d}{2}},u_{\frac{d}{2}+1}\}$ fails before $\{u_{0},u_{1}\}$, then the flow on $\{u_{0},u_{1}\}$ will be zero which means it never fails. If none of this two cases happen, then $F^{*} \setminus F_{0}=\emptyset$.
\end{proof}
\vspace*{-0.2cm}
\vspace*{-0.2cm}
\begin{claim}
\label{claim:label-D3}
For any set $F_{0} \subseteq E$, if $F^{*} \setminus F_{0} \neq \emptyset$, then $\{u_{0},u_{d+1}\} \in F_{0}$ and for any $i$, $0 \leq i \leq d$, $\{u_{i},u_{i+1}\} \notin F_{0}$.
\end{claim}
\vspace*{-0.2cm}
\vspace*{-0.2cm}
\begin{proof}
By Claim~\ref{claim:label-D2}, if $F^{*} \setminus F_{0} \neq \emptyset$, then $F^{*} \setminus F_{0} = \{u_{0},u_{1}\}$ or $F^{*} \setminus F_{0} = \{u_{\frac{d}{2}},u_{\frac{d}{2}+1}\}$. However, regarding initial flows, if $\{u_{0},u_{d+1}\} \notin F_{0}$, then $f_{u_{i},u_{i+1}}\leq m-\varepsilon\leq c_{u_{i},u_{i+1}}$ for $0\leq i\leq d$, which means that none of these lines will ever fail. Therefore, $\{u_{0},u_{d+1}\} \in F_{0}$.
\end{proof}
\vspace*{-0.2cm}
\vspace*{-0.2cm}
\begin{claim}
\label{claim:label-D4}
$D(G,|E|) \in \{0,d\}$. In particular, if $D(G,|E|) = d$, then $F^{*} \setminus F_{0} = \{u_{\frac{d}{2}},u_{\frac{d}{2}+1}\}$.
\end{claim}
\vspace*{-0.2cm}
\vspace*{-0.2cm}
\begin{proof}
Can be concluded directly from Claims~\ref{claim:label-D2} and \ref{claim:label-D3}.
\end{proof}
\vspace*{-0.2cm}
\vspace*{-0.2cm}
\begin{claim}
\label{claim:label-D5}
If there exists a solution for the instance $A = \{a_{i}, 1 \leq i \leq n\}$ of the Partition problem, then $D(G,|E|) = d$.
\end{claim}
\vspace*{-0.2cm}
\vspace*{-0.2cm}
\begin{proof}
Suppose there exists a solution for the instance $A = \{a_{i}, 1 \leq i \leq n\}$ of the Partition problem with integer values, that is there exists a subset $H \subset \{1,\dots,n\}$ such that $\sum_{i \in H} a_{i} = m$.
Let $HH = \{\{s_i,u_{0}\}|~ i \in H\}$ and $F_{0} = \{u_{0},u_{d+1}\} \cup HH$.
For any $i$, $0 \leq i \leq d$, $f_{u_{i},u_{i+1}}(F_0) = m$.
Observe that for any $i$, $0 \leq i \leq d$, $i \neq \frac{d}{2}$, $f_{u_{i},u_{i+1}}(F_0) = m\leq c_e$, but $f_{u_{\frac{d}{2}},u_{\frac{d}{2}+1}}(F_0)>c_{u_{\frac{d}{2}},u_{\frac{d}{2}+1}}$.
Therefore, $F_{1} = \{u_{\frac{d}{2}},u_{\frac{d}{2}+1}\}$ and
the system become stable because supply and demand nodes get disconnected.
Thus, $d(F_{0},F_{1}) = d$ and so $D(G,|E|) \geq d$.
By previous claims, we get $D(G,|E|) = d$.
\end{proof}
\vspace*{-0.2cm}
\vspace*{-0.2cm}
\begin{claim}
\label{claim:label-D6}
If there does not exist a solution for the instance $A = \{a_{i}, 1 \leq i \leq n\}$ of the Partition problem with integer values, then for any set of initial edge failures $F_{0} \subseteq E$, $D(G,F_{0}) = 0$.
\end{claim}
\vspace*{-0.2cm}
\vspace*{-0.2cm}
\begin{proof}
First, if $F^*\backslash F_0=\emptyset$, then $D(G,F_0)=0$ and there is nothing left to prove. Therefore, assume $F^*\backslash F_0\neq \emptyset$. Regarding Claim~\ref{claim:label-D3}, $\{u_{0},u_{d+1}\} \in F_{0}$ and for any $i$, $0 \leq i \leq d$, $\{u_{i},u_{i+1}\} \notin F_{0}$. Now assume $F_0=\{u_{0},u_{d+1}\}\cup HH$, for which $HH = \{\{s_i,u_{0}\}|~ i \in H\}$ and $H$ is an arbitrary subset of $\{1,\dots,n\}$. Since there does not exist a solution for the instance $A = \{a_{i}, 1 \leq i \leq n\}$, there are two possibilities
\begin{enumerate}
\item $\sum_{i \in H} a_{i} \leq m-1$. In this case for any $i$, $0 \leq i \leq d$, $f_{u_{i},u_{i+1}}(F_0) = m-1\leq c_{u_{i},u_{i+1}}$. Therefore, $F^*\backslash F_0 =0$ and there is nothing left to prove.
\item $\sum_{i \in H} a_{i} \geq m+1$. In this case for any $i$, $0 \leq i \leq d$, $f_{u_{i},u_{i+1}}(F_0) = m+1$. Therefore, regarding the failure model $\{u_0,u_1\}$ fails and prevent any further edge failures. After this failure, demands and supply get disconnected. Therefore, the system stabilizes. Thus $F^*=\{u_0,u_1\}\cup F_0$, and $D(G,F_0)=D(\{u_0,u_1\},F_0)=0$.
\end{enumerate}
Therefore, $D(G,F_0)=0$ in the both cases, and the proof is complete.
\end{proof}
\vspace*{-0.2cm}
From Claims~\ref{claim:label-D6} and \ref{claim:label-D5}, we can conclude that an instance $A = \{a_{i}, 1 \leq i \leq n\}$ of the Partition problem with integer values have a solution if, and only if $D(G,|E|)>0$. Since $G$ can be built from $A$ in polynomial time,  if we can approximate $D(G,|E|)$ in polynomial time by any constant, then we can check in polynomial time whether $D(G,|E|)>0$ or not. Which means that the Partition problem  can be decided in polynomial time if, and only if $D(G,|E|)$ can be approximated by a constant in polynomial time.  Now since the partition problem is well-known to be NP-hard, $D(G,|E|)$ is not in APX.
\end{proof}